\documentclass[usenatbib]{mn2e}

\usepackage{savesym}
\usepackage{amsmath}
\savesymbol{iint}
\usepackage{txfonts}
\restoresymbol{TXF}{iint}

\usepackage{color}
\usepackage[normalem]{ulem}

\usepackage{subfigure}
\usepackage{epsfig}
\usepackage{url}
\def\simless{\mathbin{\lower 3pt\hbox
{$\rlap{\raise 5pt\hbox{$\char'074$}}\mathchar"7218$}}}   
\def\simmore{\mathbin{\lower 3pt\hbox
{$\rlap{\raise 5pt\hbox{$\char'076$}}\mathchar"7218$}}}   
\newcommand{\be}{\begin{equation}}
\newcommand{\ee}{\end{equation}}
\newcommand       \bea          {\begin{eqnarray}}
\newcommand       \eea          {\end{eqnarray}}
\newcommand       \apj          {ApJ}
\newcommand       \apjl         {ApJL}
\newcommand       \aap          {A\&A}
\newcommand        \pra         {PhysRevA}
\newcommand       \nat          {Nature}
\newcommand       \mnras        {MNRAS}
\newcommand       \pasp      {PASP}

\newcommand       \prd      {Phys.~Rev.~D.~}

\newcommand      \apjs {ApJ Supplements}

\def\simlt{\mathrel{\hbox{\rlap{\hbox{\lower4pt\hbox{$\sim$}}}\hbox{$<$}}}}
\def\simgt{\mathrel{\hbox{\rlap{\hbox{\lower4pt\hbox{$\sim$}}}\hbox{$>$}}}}
\def\lesssim{\mathrel{\hbox{\rlap{\hbox{\lower4pt\hbox{$\sim$}}}\hbox{$<$}}}}

\def\gtrsim{\mathrel{\hbox{\rlap{\hbox{\lower4pt\hbox{$\sim$}}}\hbox{$>$}}}}

\title[Optical and X-ray Emission from Stable Millisecond Magnetars]{Optical and X-ray emission from stable millisecond magnetars formed from the merger of binary neutron stars}

\author[]{Brian~D.~Metzger$^{1}\thanks{E-mail: bmetzger@phys.columbia.edu}$, Anthony L.~Piro$^{2}$\\
$^{1}$Department of Physics and Columbia Astrophysics Laboratory, Columbia University, New York, NY, 10027\\
$^{2}$Theoretical Astrophysics, California Institute of Technology, 1200 E California Blvd., M/C 350-17, Pasadena, CA 91125}

\begin{document}
\date{Received / Accepted}
\pagerange{\pageref{firstpage}--\pageref{lastpage}} \pubyear{2012}

\maketitle

\label{firstpage}

\begin{abstract}

The coalescence of binary neutron stars (NSs) may in some cases produce a stable massive NS remnant rather than a black hole. Due to the substantial angular momentum from the binary, such a remnant is born rapidly rotating and likely acquires a strong magnetic field (a `millisecond magnetar'). Magnetic spin-down deposits a large fraction of the rotational energy from the magnetar behind the small quantity of mass ejected during the merger. This has the potential for creating a bright transient that could be useful for determining whether a NS or black hole was formed in the merger. We investigate the expected signature of such an event, including for the first time the important impact of  $e^{\pm}$ pairs injected by the millisecond magnetar into the surrounding nebula. These pairs cool via synchrotron and inverse Compton emission, producing a pair cascade and hard X-ray spectrum.  A fraction of these X-rays are absorbed by the ejecta walls and re-emitted as thermal radiation, leading to an optical/UV transient peaking at a luminosity of $\sim 10^{43}-10^{44}\,{\rm erg \,s^{-1}}$ on a timescale of several hours to days.  This is dimmer than predicted by simpler analytic models because the large optical depth of $e^{\pm}$ pairs across the nebula suppresses the efficiency with which the magnetar spin down luminosity is thermalized.  Nevertheless, the optical/UV emission is more than two orders of magnitude brighter than a radioactively powered `kilonova.'  In some cases nebular X-rays are sufficiently luminous to re-ionize the ejecta, in which case non-thermal X-rays escape the ejecta unattenuated with a similar peak luminosity and timescale as the optical radiation.  We discuss the implications of our results for the temporally extended X-ray emission that is observed to follow some short gamma-ray bursts (GRBs), including the kilonova candidates GRB 080503 and GRB 130603B.
\end{abstract} 
  

\section{Introduction} 
\label{intro}

The coalescence of binary neutron stars (NSs) are the leading model for generating the central engines that power short-duration gamma-ray bursts (GRBs) (\citealt{Paczynski86}; \citealt{Eichler+89}). In addition, NS mergers are expected to be the primary source of gravitational waves for upcoming ground-based interferometric detectors such as Advanced LIGO and Virgo (\citealt{Abadie+10}).  It is not known whether a relativistic jet as needed for a GRB is generated in all NS mergers, and even if there was, jets that are beamed away from the observer would be missed. Therefore, understanding the full range of potential electromagnetic counterparts to such gravitational wave events is crucial for localization and building a complete physical picture of what occurs in these mergers.

An important factor in determining the electromagnetic signature is the properties of the remnant left over from the merger.  Numerical simulations of the merger process (e.g.,~\citealt{ruffert1999}; \citealt{Uryu+00}; \citealt{Rosswog&Liebendorfer03}; \citealt{Oechslin&Janka06}; \citealt{Chawla+10}; \citealt{Rezzolla+10};
\citealt{Hotokezaka+11}) show that initially the end product is a hypermassive NS, which is (at least temporarily) stable to gravitational collapse as a result of support by thermal pressure and/or differential rotation (\citealt{Morrison+04}; \citealt{OConnor&Ott11}; \citealt{Paschalidis+12}; \citealt{Lehner+12}; \citealt{Kaplan+13}). It has generally been assumed that this NS subsequently collapses to form a black hole within a relatively short timescale $\lesssim 100$ ms.  The newly created black hole is surrounded by a thick remnant torus of mass $M_{\rm t} \sim 10^{-2} M_{\odot}$, which accretes and powers the transient relativistic jet responsible for the GRB (e.g.,~\citealt{Narayan+92}; \citealt{Rezzolla+11}).

Another possibility is that rather than a black hole,  the merger results in a {\it massive NS that is indefinitely stable to gravitational collapse.} A major uncertainty in understanding whether this occurs depends on our incomplete knowledge of the equation of state of high density matter \citep{Hebeler+13}.  The recent discovery of massive $\sim 2M_{\odot}$ NSs (\citealt{Demorest+10}; \citealt{Antoniadis+13}) indicates that the EoS is not as soft as predicted by some previous models.  This makes it more likely that a massive NS remnant could remain (e.g.,~\citealt{Ozel+10}; \citealt{Bucciantini+12}; \citealt{Giacomazzo&Perna13}; \citealt{Kiziltan+13}).  Due to the large angular momentum of the merging binary, such a NS will necessarily be rotating extremely rapidly, with a rotation period $P \sim 1\,{\rm ms}$, close to the centrifugal break-up limit.  The NS remnant may also acquire a strong magnetic field $B_{\rm d} \gtrsim 10^{14}\,{\rm G}$ similar to `magnetars' either due to the amplification of an initially weak field by shear instabilities at the merger interface (\citealt{Price&Rosswog06}; \citealt{Zrake&MacFadyen13}) or by an $\alpha-\Omega$ dynamo (\citealt{Duncan&Thompson92}) in the subsequent neutrino cooling phase.

Magnetic spin-down of a NS remnant represents an additional source of sustained energy injection which is not present in the case of prompt black hole formation.  Such remnants have been suggested as a possible explanation for the X-ray activity observed after some short GRBs, such as the temporally `extended emission' observed on timescales of minutes after the GRB (\citealt{Gao&Fan06}; \citealt{Metzger+08}; \citealt{Metzger+11}; \citealt{Bucciantini+12}) and the X-ray `plateaus' observed on longer timescales $\sim 10^{2}-10^{4}$ s (\citealt{Rowlinson+13}; \citealt{Gompertz+13}).  In contrast, accretion of the remnant torus cannot readily explain such emission, since the torus is not present at such late times due to powerful outflows that occur on timescales of seconds or less (e.g.,~\citealt{Metzger+08}; \citealt{Lee+09}; \citealt{Dessart+09}; \citealt{Fernandez&Metzger13}). Prolonged energy injection from a magnetar remnant has also been proposed as a potential electromagnetic counterpart to the gravitational wave signal (\citealt{Zhang13}; \citealt{Gao+13}; \citealt{Yu+13}).  



The region surrounding the site of the merger is polluted by mass ejected dynamically during the merger itself (e.g.,~\citealt{Rosswog+12}; \citealt{Hotokezaka+13}), and in subsequent outflows from the remnant accretion disk or the hyper-massive NS (\citealt{Metzger+08}; \citealt{Dessart+09}; \citealt{Caballero+12}; \citealt{Fernandez&Metzger13}).  At early times, and for magnetars with a sufficiently large spin-down luminosity, the outflow from the magnetar wind may be collimated into a bipolar jet by its interaction with this ejecta (\citealt{Bucciantini+12}), which could produce a phase of extended high energy emission similar to a GRB (\citealt{Metzger+08}).  However, at later times as the spin-down luminosity decreases $L_{\rm sd} \propto t^{-2}$ , it becomes increasingly likely that such a jet will be stifled behind the ejecta, due to the growth of MHD instabilities and resulting magnetic reconnection (\citealt{Porth+13}).  

To estimate when a jet is expected to be stifled, note that the time required for a jet of luminosity $L_j$ and (fixed) opening angle $\theta$ to drill through an envelope of radius $R_{\rm ej}$ and mass $M_{\rm ej}$ is given by \citep{Bromberg+11}
\be
t_{\rm bo} \approx 0.17\left(\frac{L_{\rm j}}{10^{47}{\rm erg\,s^{-1}}}\right)^{-1/3}\left(\frac{\theta}{30^{\circ}}\right)^{4/3}\left(\frac{R_{\rm ej}}{10^{13}{\rm cm}}\right)^{2/3}\left(\frac{M_{\rm ej}}{0.01M_{\odot}}\right)^{1/3}\,{\rm hr}\,
\ee  
If we apply this to the expanding merger ejecta of radius $R_{\rm ej} \simeq v_{\rm ej} t$ and velocity $v_{\rm ej}$, and if we assume that the jet luminosity contains a fraction $\epsilon_{\rm j}$ of the pulsar spin-down luminosity $L_{\rm sd} \simeq 5\times 10^{47}(B_{\rm d}/10^{15}{\rm G})^{-2}(t/{\rm hr})^{-2}$ (calculated at times much greater than the initial spin-down time; eq.~[\ref{eq:Lsd}]), then we find that
\be
\frac{t_{\rm bo}}{t} \sim 0.6 \left(\frac{\epsilon_{j}}{0.5}\right)^{-1/3}\left(\frac{B_{\rm d}}{10^{15}{\rm G}}\right)^{2/3}\left(\frac{\theta}{30^{\circ}}\right)^{4/3}\left(\frac{v_{\rm ej}}{c}\right)^{2/3}\left(\frac{M_{\rm ej}}{0.01M_{\odot}}\right)^{1/3}\left(\frac{t}{\rm hr}\right)^{1/3},
\label{eq:jbo}
\ee
where the jet efficiency $\epsilon_{j}$ is normalized to a relatively high value.  Thus, depending on the value of $B_{\rm d}$ and the jet opening angle $\theta$, the time for the jet to escape the ejecta becomes longer than the evolution timescale $t$ on a timescale of hours.  When $t_{\rm bo} \gtrsim t$ the jet may no longer be able to propagate steadily through the ejecta.

Even if the magnetar does not produce high energy jetted radiation directly, its rotational energy is deposited into a hot nebula behind the expanding ejecta.  Emission from this nebula will escape to the observer once the ejecta has expanded sufficiently that photons can diffuse outwards on the expansion timescale, as occurs on timescales of several hours to days. Such sources of emission are of great interest as a possible electromagnetic counterpart to the gravitational waves produced by the inspiral chirp. Past work on `kilonovae' has focused on emission powered solely by the radioactive decay of heavy nuclei synthesized in the ejecta (\citealt{Li&Paczynski98}; \citealt{Metzger+08}; \citealt{Metzger+10}; \citealt{Roberts+11}; \citealt{Piran+13} \citealt{Tanaka&Hotokezaka13}; \citealt{Tanaka+13}), as is arguably the most promising counterpart for the majority of mergers leaving black hole remnants (\citealt{Metzger&Berger12}).  The rotational energy injected by a stable millisecond magnetar ($\sim 10^{52}$ ergs) is, however, significantly greater than that produced by radioactivity ($\lesssim 10^{47}$ ergs on a timescale of $\sim$ days), potentially producing a much more luminous signal. Such emission would be a {\it smoking gun for a long-lived magnetar remnant,} and its  detection would place important constraints on the equation of state of nuclear density matter.


In this paper we calculate the evolution of a millisecond magnetar nebula produced by a NS merger and its associated thermal (optical/UV) and non-thermal (X-ray) emission. Our basic framework is an extension of a model developed recently by \citet{Metzger+13} for application to the different physical context of superluminous supernovae. During the preparation of our work, \citet{Yu+13} also investigated the potential thermal signature of a millisecond magnetar remnant. Although the physical situation we consider is similar, there are notable differences between our predictions and those of \citet{Yu+13}.  In particular, the optical luminosities that we calculate are significantly lower. As we summarize in the next section, this difference is due to their neglect of the high optical depth of nebular pairs which impacts the efficiency with which the pulsar wind thermalizes. Previous works also ignore the important role of ionization on the X-ray opacity of the ejecta and its resulting non-thermal radiation.

This paper is organized as follows.  In \S\ref{sec:basic} we provide an overview the basic physical model.  In \S\ref{sec:model} we describe the model details and in \S\ref{sec:evo} we summarize the equations governing the coupled evolution of thermal and non-thermal radiation in the  magnetar nebula and the resulting radiation.  In \S\ref{sec:results} we present our results, which we justify based on a simple physical model in Appendices \ref{sec:appendixLC} and \ref{sec:ion}.  We discuss our results in \S\ref{sec:discussion} and summarize our conclusions in \S\ref{sec:conclusions}.

\section{Basic Physical Model} 
\label{sec:basic}

\begin{figure*}
\includegraphics[width=1.0\textwidth]{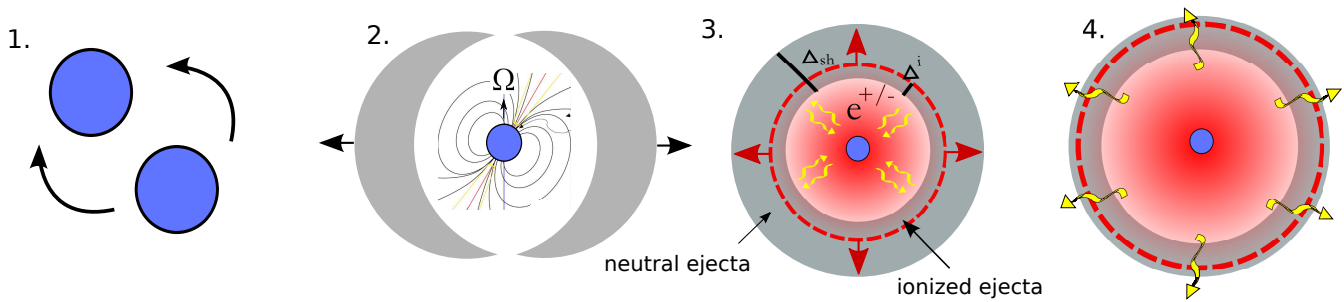}
\caption{Stages for the optical/X-ray emission from a binary NS merger that results in a stable millisecond magnetar remnant.  (1) NS binary merges due to the emission of gravitational radiation. (2) The merger produces a stable millisecond magnetar.  Following the merger, a mass $M_{\rm ej} \sim 0.01-0.1\,M_{\odot}$ is ejected with an initial velocity $v_{\rm ej} \sim 0.1$ c, encompassing the magnetar in a quasi-spherical shell.  (3) Magnetar wind dissipates its energy behind the expanding ejecta, producing a nebula of e$^{\pm}$ pairs and non-thermal photons.  Non-thermal UV and X-ray photons thermalize via their interaction with the ejecta walls.  This thermalization process is in some cases suppressed due to the large pair optical depth through the nebula, which decreases the effective rate that non-thermal photons transfer their energy to the surrounding ejecta.  Nebular X-rays drive an ionization front ({\it red dashed line}) outwards through the ejecta.  (4) On a timescale $\sim t_{\rm d,0}^{\rm ej} \sim$ several hours to days, optical and X-ray photons diffuse out of the nebula, producing luminous emission.
}
\label{fig:schematic}
\end{figure*}

The basic physical picture of our model is summarized in Figure~\ref{fig:schematic} and described below. This sets the stage for the more detailed investigation that we present for the rest of the paper.

Following binary coalescence (Stage 1), a massive NS is formed with a millisecond rotation period and a strong magnetic dipole field $B_{\rm d} \gtrsim 10^{14}\,{\rm G}$.  Initially the magnetar is surrounded by a centrifugally-supported accretion disk, but most of this disk either accretes or is unbound by outflows on a relatively short timescale $\lesssim$ seconds (\citealt{Metzger+08b}; \citealt{Dessart+09}; \citealt{Lee+09}; \citealt{Fernandez&Metzger13}).  Within a few seconds following the merger, the magnetar is thus surrounded by an expanding shell of ejecta with mass $M_{\rm ej} \sim 0.01-0.1\,M_{\odot}$ and initial velocity $v \sim 0.1c$ (Stage 2).

Assuming that the inertia of the ejecta is sufficient to stifle the formation of a jet (eq.~[\ref{eq:jbo}]), the energy of the magnetar wind is dissipated behind the ejecta via shocks or magnetic reconnection (as is observed to occur in normal pulsar wind nebulae; \citealt{Kennel&Coroniti84}), forming a hot nebula of photons and electron/positron pairs (Stage 3).  At early times this nebula is small and opaque.  Most injected energy is lost to PdV work, causing the ejecta to accelerate to velocities $v \sim c$ on a timescale comparable to the pulsar spin-down time $\sim$ minutes$-$hours.  In such an environment newly-injected pairs rapidly lose their energy to inverse Compton scattering and synchrotron radiation.  This hard radiation in turn produces more pairs, resulting in a `cascade' of pair production and a non-thermal X-ray spectrum $J_{\nu} \propto \nu^{-\alpha}$ with $\alpha \approx 1$ (\citealt{Svensson87}).  

Nebular X-rays that encounter the ejecta walls are either reflected or absorbed, depending on the albedo of the ionized layer that separates the nebula from the outer neutral ejecta.  The albedo in turn depends on the ionization state of the layer, which itself depends on the irradiating X-ray flux.  The efficiency with which the pulsar X-rays are thermalized depends not only on the absorption efficiency of the ejecta, {\it but also on the very presence of pairs in the nebula}.  When the pair optical depth of the nebula is high, then X-rays from the pulsar lose a substantial fraction of their energy to PdV work before they can cross the nebula to be thermalized at the ejecta walls. As we will show, the effects of (1) a realistic ejecta albedo and (2) a high pair optical depth, can substantially reduce the optical luminosity in comparison to models that assume the pulsar luminosity is thermalized with high efficiency.

As the ejecta expands, its optical depth decreases.  Once the photon diffusion timescale becomes shorter than the expansion timescale, photons are able to escape the nebula. At optical wavelengths where the opacity is relatively low, this transition occurs on a timescale of several hours to days.  At X-ray wavelengths, however, the bound-free opacity remains orders of magnitude higher because the ejecta recombines to become neutral as it expands and cools.  If the ejecta remains neutral X-rays are effectively trapped on time-scales of interest.  However in many cases the nebular X-rays are sufficiently luminous to re-ionize the ejecta, after which time the X-ray opacity is instead set by the much lower value due to Thomson scattering.  Ionization, if it occurs, thus allows X-rays also to escape the nebula on a time-scale which is is typically comparable to the time of peak optical emission.  At later times, as the ionized ejecta becomes optically thin, X-rays are thermalized with decreasing efficiency.  Emission thus becomes dominated by the non-thermal radiation, with its total luminosity tracking the decaying pulsar spin-down power $\propto t^{-2}$.

\section{Model}
\label{sec:model}

In this section we describe our model for the evolution of millisecond magnetar wind nebulae following binary NS mergers.  The model is similar to that developed by \citet{Metzger+13} for superluminous supernovae, but is applied here to different physical conditions corresponding to a much lower ejecta mass, higher ejecta velocity, and a NS with a stronger magnetic field. This provides the basic ingredients that will be used in the evolution equations we present in the \S\ref{sec:evo}.

\subsection{Merger Ejecta}
\label{sec:ejecta}

The merger site is surrounded by an envelope of mass $M_{\rm ej} = 10^{-2}M_{-2} M_{\odot}$ moving outwards with an initial velocity $v_{\rm ej,0} = c\beta_{0} \sim 0.1$ $c$ and density profile
\be
\rho_{\rm ej}(r,t) = \frac{3}{4\pi}\frac{M_{\rm ej}}{R(t)^{3}},
\label{eq:rho0}
\ee
where $R$ is the characteristic radius of the ejecta.  A uniform density profile is assumed for simplicity since our results are not particularly sensitive to the precise radial mass distribution.  

The ejecta includes both matter which is unbound dynamically during the merger itself (`tidal tails') and that produced subsequently by outflows from the accretion disk or remnant NS (\citealt{Dessart+09}; \citealt{Metzger+12}).  The tidal tails are concentrated primarily in the plane of the original orbit, although the effects of self-similar expansion (\citealt{Roberts+11}) and radioactive heating (\citealt{Metzger+10b}; \citealt{Rosswog+13b}) tend to smooth out the density distribution.  Outflows from the accretion disk and the remnant NS are more spherically symmetric, though are still equatorially concentrated to some extent (\citealt{Fernandez&Metzger13}).  At a minimum the magnetar will be surrounded by the mass ejected by the neutrino-driven wind from its surface, which is $\gtrsim 10^{-3}M_{\odot}$, depending on the strength of the magnetic field and neutrino cooling evolution of the NS remnant (e.g.~\citealt{Metzger+07}).   

Our model assumes spherically symmetric ejecta since we are primarily focused on the interaction between the magnetar wind and the disk wind material. The geometry of the ejecta and nebula is illustrated in Stage 3 of Figure \ref{fig:schematic}. We also assume that it is composed of iron-rich elements, as results from the higher electron fraction produced by neutrino irradiation from the remnant NS (\citealt{Metzger+09a}; Fernandez et al., in prep). This is in contrast to the tail material, which is highly neutron-rich and is composed of heavy $r$-process elements with $A \gtrsim 130$ (e.g.,~\citealt{Freiburghaus+99}).


Thermal radiation is free to diffuse out of the ejecta once the photon diffusion timescale $t_{\rm d}^{\rm ej} \sim 3M\kappa/4\pi Rc$ becomes less than the expansion timescale of the ejecta $\sim R/\beta c$, where $\kappa$ is the opacity (Arnett 1982).  This occurs after a characteristic time
\be
t_{\rm d,0}^{\rm ej} = \left(\frac{3}{4\pi}\frac{\kappa M}{vc}\right)^{1/2} \approx 9M_{-2}^{1/2}(\kappa/\kappa_{\rm es})^{1/2}\beta^{-1/2}\,{\rm hr}.
\label{eq:td}
\ee
Hereafter we assume a constant opacity $\kappa = \kappa_{es} \approx 0.2\,{\rm cm^{2}\,g^{-1}}$ at optical/UV frequencies, as approximately characterizes the line opacity of Fe (\citealt{Pinto&Eastman00}). Recent work by \citet{Kasen+13} has shown that the opacity of $r$-process material is orders of magnitude higher than $\kappa_{\rm es}$ due the bound-bound opacities of Lanthanide elements (\citealt{Barnes&Kasen13}), but we do not use this since we are focused on disk wind material.  Furthermore, as we show in \S\ref{sec:lanthanides}, X-rays from the magnetar nebula may be sufficiently luminous to ionize the valence electrons of the Lanthanides, in which case their opacity would be substantially reduced anyway.

The optical depth of the ejecta at optical frequencies decreases with time as
\be
\tau^{\rm ej}_{\rm es} \simeq \frac{3}{4\pi}\frac{M\kappa}{R^{2}} \approx \beta^{-1}\left(\frac{t}{t_{\rm d,0}^{\rm ej}}\right)^{-2}
\label{eq:tau_ej},
\ee
such that after a time
\be
t_{\rm thin}^{\rm ej} = t_{\rm d,0}^{\rm ej}\beta^{-1/2} \approx 9M_{-2}^{1/2}\beta^{-1}\,{\rm hr},
\label{eq:tthinej}
\ee
the ejecta becomes optically thin ($\tau_{\rm es}^{\rm ej} < 1$).

\subsection{Pulsar Wind Nebula}

A NS with an initial rotational period $P = 2\pi/\Omega = P_{-3}\,{\rm ms}$ has an associated rotational energy
\be
E_{\rm rot} = \frac{1}{2}I\Omega^{2} \simeq 3\times 10^{52}P_{-3}^{-2}\,{\rm erg},
\label{eq:Erot}
\ee 
where $I \simeq 2\times 10^{45}\,{\rm g\, cm^{2}}$ is the NS moment of inertia.  An initial period $P_{-3} \sim 1$ close to the centrifugal break-up limit is expected due to the substantial angular momentum of the initial binary (e.g.,~\citealt{Giacomazzo&Perna13}).   Spin-down is enhanced during the first $\sim$ minute due to the enhanced torque of the neutrino-driven wind (\citealt{Thompson+04}), which detailed calculations show acts to increase the period to $\sim 2\,{\rm ms}$ (Metzger et al. 2011).  Very rapid rotation may also lead to self-gravitational instabilities and resulting gravitational wave radiation (e.g.,~\citealt{Piro&Ott11}).  We thus adopt $P = 2\,{\rm ms}$ as a characteristic `initial' period, corresponding to $E_{\rm rot} \approx 10^{52}\,{\rm erg}$. 

The pulsar injects energy behind the ejecta at a rate, which for an aligned force-free wind is given by
\begin{eqnarray}
L_{\rm sd} = \frac{\mu^{2}\Omega^{4}}{c^{3}} \simeq 6\times 10^{49}B_{15}^{2}P_{-3}^{-4}\left(1 + \frac{t}{t_{\rm sd}}\right)^{-2}{\rm erg\,s^{-1}} \nonumber \\
\underset{t \gg t_{\rm sd}}\approx  1.2\times 10^{48}B_{15}^{-2}t_{\rm hr}^{-2}\,{\rm erg\,s^{-1}}
\label{eq:Lsd}
\end{eqnarray}
where $t \equiv t_{\rm hr}\,{\rm hr}$ is time since the merger, $\mu = B_{\rm d}R_{\rm NS}^{3}$ is the dipole moment, $B_{\rm d} = 10^{15}B_{15}\,{\rm G}$ is the surface equatorial dipole field, $R_{\rm NS} = 10\,{\rm km}$ is the NS radius, and
\be
t_{\rm sd} = \left.\frac{E_{\rm rot}}{L_{\rm sd}}\right|_{t = 0}\simeq 0.14B_{15}^{-2}P_{-3}^{2}{\rm \,hr\,}
\label{eq:tsd}
\ee
is the initial spin-down time.  This luminosity inflates a nebula of relativistic particles and radiation inside the cavity evacuated by the expanding merger ejecta (Figure \ref{fig:schematic}).  Note that $t_{\rm sd}$ is less than the characteristic diffusion time $t_{\rm d,0}^{\rm ej}$ (eq.~[\ref{eq:td}]) for magnetar-strength fields $B_{\rm d} \gtrsim 10^{14}\,{\rm G}$ of interest.

Pulsar winds are composed primarily of $e^{\pm}$ pairs, which are heated to a very high energy per particle (random Lorentz factor $\gtrsim 10^{4}$) by the shock or reconnection layer responsible for dissipating the bulk of the wind energy (\citealt{Metzger+13}).  The extremely high energy density of radiation and magnetic fields in the nebula results in pairs rapidly cooling on timescales much less than the evolution timescale via synchrotron emission and inverse Compton scattering.  The photons so produced are typically of sufficient energy $\gg m_e c^{2}$ to produce additional $e^{\pm}$ pairs.  Such pair creation is copious due to the high compactness parameter of the nebula,
\begin{eqnarray}
\ell &\equiv& \frac{E_{\rm nth}\sigma_{\rm T} R}{Vm_e c^{2}} \underset{t \gg t_{\rm sd}}\sim 3\times 10^{4}B_{15}^{-2}t_{\rm hr}^{-3}
\label{eq:compactness}
\end{eqnarray}
where $V = 4\pi R^{3}/3$ is the volume of the nebula and in the numerical estimate the radius is approximated as $R \sim ct$ and the non-thermal energy of the nebula $E_{\rm nth}$ as $\sim L_{\rm sd}t$.\footnote{This latter approximation is only accurate at times much less than the characteristic diffusion time of the nebula $t_{\rm d,0}^{\rm n}$ (see eq.~[\ref{eq:tdn}] below).}  The pair cascade thus resides in a so-called `saturated' state ($\ell \gg 1$; e.g.,~\citealt{Svensson87}) until times $\gtrsim t_{\rm d,0}^{\rm ej} \sim t_{\rm thin}^{\rm ej}$ for magnetars with $B_{\rm d} \lesssim$ few $\times 10^{15}$ G.

The total number of pairs $N_{\pm}$ in the nebula evolves according to 
\be
\frac{dN_{\pm}}{dt} = \dot{N}_{\pm}^{+} - \dot{N}_{\pm}^{-},
\label{eq:Npmevo}
\ee
where
\be
\dot{N}_{\pm}^{\rm +} \simeq \frac{Y L_{\rm sd}}{m_e c^{2}} 
\label{eq:Ndotplus}
\ee
is the pair creation rate, $Y \simeq 0.1$ is the pair multiplicity for a saturated cascade (\citealt{Svensson87}), and
\be
\dot{N}_{\pm}^{\rm -} = \frac{3}{16}\sigma_{\rm T} c N_{\pm}^{2}V^{-1},
\label{eq:Ndotminus}
\ee
is the pair annihilation rate.

At early times, the optical depth of pairs through the nebula is determined by the balance between creation $\dot{N}_{\pm}^{+}$ (eq.~[\ref{eq:Ndotplus}]) and destruction $\dot{N}_{\pm}^{-}$ (eq.~[\ref{eq:Ndotminus}]), giving
\begin{eqnarray}
\tau_{\rm es}^{\rm n} &=& \sigma_{\rm T} Rn_{\pm} = R\left[\frac{16 \dot{N}_{\pm}^{-}\sigma_{\rm T}}{3  V c}\right]^{1/2} 
= \left[\frac{4 Y\sigma_{\rm T} L_{\rm sd}}{\pi R m_e c^{3}}\right]^{1/2} \nonumber \\ 
&\simeq& \left\{
\begin{array}{lr}
1.3\times 10^{3}\,\,B_{15}\beta^{-1/2}P_{-3}^{-1}t_{\rm hr}^{-1/2}
, 
&(t \ll t_{\rm sd} ) \\
120\,\, B_{15}^{-1}\beta^{-1/2}t_{\rm hr}^{-3/2}       
,&(t \gg t_{\rm sd}), \\
\end{array}
\right.
\label{eq:taueseq}
\end{eqnarray}
The timescale to reach this equilibrium, $t_{\rm eq} \simeq 16 R/3 c \tau_{\rm es}^{\rm n}$, is short compared to the evolution timescale $t = R/v$ as long as
\be
\tau_{\rm es}^{\rm n} \gg \frac{16}{3}\frac{v}{c} \simeq 5.3\beta
\ee
Since $\tau_{\rm es}^{\rm n}$ is only relevant insofar as it is $\gtrsim 1$, equation (\ref{eq:taueseq}) is an excellent approximation for practical purposes.  Note that the pair optical depth $\tau_{\rm es}^{\rm n}$ {\it often exceeds that of the ejecta itself} $\tau_{\rm es}^{\rm ej}$ (eq.~[\ref{eq:tau_ej}]).  This crucial fact will prove relevant later in calculating the efficiency with which the pulsar luminosity can be radiated as optical emission.  Also note that $\tau_{\rm es}^{\rm n} \sim \ell^{1/2}$ for $v_{\rm ej} \sim c$; thus a saturated pair cascade ($\ell \gg 1$) is necessarily maintained while $\tau_{\rm es}^{\rm n} \gtrsim 1$. 
    
Radiation travels freely across the nebula once the photon diffusion timescale across the nebula 
\be t_{\rm d}^{\rm n} \simeq \frac{R}{c}(\tau_{\rm es}^{\rm n}+1)
\label{eq:tdiffn}
\ee
 becomes less than the expansion timescale $\sim R/\beta c$.  This occurs after a characteristic time
\be
t_{\rm d,0}^{\rm n} = 24\,B_{15}^{-2/3}\beta^{1/3}\,{\rm hr},
\label{eq:tdn}
\ee
 The factor of unity inside the parentheses in equation (\ref{eq:tdiffn}) extrapolates the cooling timescale smoothly from the optically-thick to optically-thin cases; in the latter case $t_{\rm d}^{\rm n}$ becomes the light crossing time.  The nebula becomes thin to pairs ($\tau_{\rm es}^{\rm n} < 1$) after a time
\be
t_{\rm thin}^{\rm n} = \beta^{-2/3}t_{\rm d,0}^{\rm n} = 24\,B_{15}^{-2/3}\beta^{-1/3}\,{\rm hr}.
\label{eq:tthinn}
\ee
Note the close analogy between equations (\ref{eq:tdn}) and (\ref{eq:tthinn}) for the timescale for photon propagation through the nebula, and those in equations (\ref{eq:td}) and (\ref{eq:tthinej}) for propagation out through the ejecta.  These timescales differ because, although the number of electrons in the ejecta is fixed, the number of $e^{\pm}$ pairs in the nebula changes continually due to the balance between creation and annihilation (eq.~[\ref{eq:taueseq}]). As we will subsequently show, it is the interplay between these different timescales that will contribute to the lower efficiency in producing thermal emission from the heated ejecta as well as imprint a unique time evolution for the thermal luminosity (also see Appendix \ref{sec:appendixLC}).


\section{Evolution Equations}
\label{sec:evo}

We now summarize the equations describing the coupled evolution of the nebula and the ejecta.  Part of our model is based on that of \citet{Metzger+13}, so the reader is encouraged to consult this reference for further details and clarification.

\subsection{Radius}
The ejecta and the nebula are assumed to share a common radius $R$.  This is a good approximation following a short-lived phase at early times during which the nebula drives a shock through the ejecta.  This common radius evolves according to
\be
\frac{dR}{dt} = \beta c = \left(\frac{2\int_0^{t} L_{\rm sd}dt'}{M_{\rm ej}} + v_{\rm ej,0}^{2}\right)^{1/2},
\label{eq:dRdt}
\ee
where $\int_0^{t}L_{\rm sd}dt'$ is the rotational energy injected by the pulsar up to the time of interest.  Energy conservation is justified since the majority of the non-thermal energy injected into the nebula by the pulsar is lost to PdV work instead of being radiated (see below).  At times $t \gg t_{\rm sd}$, the radius thus evolves as $R \simeq \beta c t$, where $\beta \sim 1.0(E_{\rm rot}/10^{52}\,{\rm erg})^{1/2}M_{-2}^{-1/2}$ is the asymptotic velocity.  Relativistic corrections are neglected for simplicity, since our other simplifications result in a comparable or greater loss of accuracy.

\subsection{Nebular Radiation}
The photon spectrum of the nebula is comprised of a thermal bath and a non-thermal tail.  The spectrum of non-thermal radiation $E_{\rm nth,\nu}$ per unit frequency evolves according to
\be
\frac{dE_{\rm nth,\nu}}{dt} =  -\frac{E_{\rm nth,\nu}}{R}\frac{dR}{dt} + \dot{E}_{\rm sd,\nu} - L_{\rm nth,\nu} - (1-\mathcal{A}_{\nu})\frac{E_{\rm nth,\nu}}{t_{\rm d}^{\rm n}} ,
\label{eq:photon_evo}
\ee
while the thermal energy of the ejecta evolves according to
\be
\frac{dE_{\rm th}}{dt} = -\frac{E_{\rm th}}{R}\frac{dR}{dt}  - L_{\rm th} + \int (1-\mathcal{A}^{\rm th}_{\nu})\frac{E_{\rm nth,\nu}}{t_{\rm d}^{\rm n}} d\nu.
\label{eq:thermo_evo}
\ee
The first term in both equations accounts for PdV losses.\footnote{The first term in equation (\ref{eq:photon_evo}) does not properly take into account the {\it redistribution} of photon energy due to adiabatic expansion.  Nevertheless, this expression is valid for a flat spectrum with $\nu E_{\rm nth, \nu} \sim$ constant, as approximately characterizes the nebular spectrum in our calculations.}  The second term in equation (\ref{eq:photon_evo})   ,
\be
\dot{E}_{\rm sd,\nu} \simeq \frac{L_{\rm sd}}{14\nu}\,\,\,\,\,\,\,\,(3kT_{\rm th}  \sim 1 \,\,{\rm eV} \lesssim h\nu \lesssim 1 \,\,{\rm MeV})
\label{eq:edotsd}
\ee
accounts for the photon spectrum injected by the pair cascade, normalized such that $\int \dot{E}_{\rm sd,\nu}d\nu$ equals the total pulsar power $L_{\rm sd}(t)$ (eq.~[\ref{eq:Lsd}]).  Here we have assumed that injected photon spectrum extends from the thermal bath energy $\sim 3kT_{\rm th}$ of temperature $T_{\rm th} = (E_{\rm th}/a V)^{1/4}$ up to the pair creation threshold $\sim 2 m_e c^{2} \sim$ MeV.  The second term in equation (\ref{eq:thermo_evo}), 
\be
L_{\rm th} = \frac{E_{\rm th}}{t_{\rm d}^{\rm ej}},
\label{eq:L_SN}
\ee
accounts for radiative losses from the thermal bath, as is responsible for powering the optical luminosity, where
\be
t_{\rm d}^{\rm ej} =  \left(\tau_{\rm es}^{\rm ej} + 1\right)\frac{R}{c},
\label{eq:td2}
\ee
is the characteristic photon diffusion timescale through the ejecta for thermal photons, where $\tau_{\rm es}^{\rm ej}$ is the optical depth of the ejecta (eq.~[\ref{eq:tau_ej}]).


The last terms in equations (\ref{eq:photon_evo}) and (\ref{eq:thermo_evo}) account for the absorption of photons by the ionized inner layer of the ejecta, where $t_{\rm d}^{\rm n}$ (eq.~[\ref{eq:tdiffn}]) is the timescale for photons to propagate across the nebula.  This equals the characteristic interval between each interaction of a nebular photon with the ejecta, where the pair optical depth $\tau_{\rm es}^{\rm n} = N_{\pm}\sigma_T R$ is calculated by simultaneously evolving the pair number $N_{\pm}$ according to equation (\ref{eq:Npmevo}).


The factor $\mathcal{A}_{\nu}$ is the albedo of the ejecta walls to a photon of frequency $\nu$, i.e. the probability of reflection.  The factor $1 - \mathcal{A}_{\nu}^{\rm th}$ is the probability that a photon is absorbed, or loses significant energy to down-scattering.  All energy absorbed by the ejecta is assumed to be thermalized.  This is a strong assumption, since a fraction of the radiation could also be radiated at the Compton temperature of the electrons in the ionized layer, which in general can be significantly higher than the thermal blackbody temperature of the photons. Given that the present model we are using already has many interesting features to explore, and our treatment of the physics is significantly more detailed than other similar investigations into such systems, we feel that making this assumption is reasonable for a first step.

The albedo $\mathcal{A}_{\nu}$ and the absorption efficiency $\mathcal{A}_{\nu}^{\rm th}$ depend on the (frequency-dependent) ratio of the absorption and scattering opacities\footnote{Note that although we use the subscript $\nu$ for some opacities, this merely indicates a frequency dependence for these opacities and the units are still ${\rm cm^2\,g^{-1}}$ in cgs.}
\be \eta \equiv \frac{\kappa_{\rm abs,\nu}}{\kappa_{\rm es}},
\label{eq:eta1}
\ee 
in the ionized layer of the atomic species $i$ which dominates the penetration depth of photons with frequencies in the range of interest,
where 
\be 
\kappa_{\rm abs,\nu} = \kappa_{\rm bf,\nu}^{i} + \kappa_{\rm ie},
\ee
$\kappa_{\rm bf,\nu}^{i}$ is the bound-free opacity (eq.~[\ref{eq:kappa}]) of species $i$, and
\be
\kappa_{\rm ie} = \kappa_{\rm es}\frac{h\nu}{m_{\rm e}c^{2}}
\ee
is the effective absorptive opacity for a photon of frequency $\nu$ due to inelastic down-scattering.  

As in \citet{Metzger+13}, we calculate $\mathcal{A}_{\nu}$ and $\mathcal{A}_{\nu}^{\rm th}$ using a Monte Carlo procedure by which the fate of a large number of photons injected into one side of a slab (representing the ionized layer) is followed, counting the fraction absorbed, reflected, or transmitted through the slab.  The thickness of the ionized layer $\Delta^{i}$ used in performing this calculation depends on whether the ejecta is completely ionized ($\Delta^{i} = R$), or whether the outer layers remain neutral ($\Delta^{i} < R$). In the latter case, any photon which is not absorbed in the ionized layer must be absorbed by the ejecta further out, such that $\mathcal{A}_{\nu}^{\rm th} = \mathcal{A}_{\nu}$.  For a given value of $\eta$, the thickness of the ionized layer used in calculating $\mathcal{A}_{\nu}^{\rm th} = \mathcal{A}_{\nu}$ in this case is determined by the condition that the effective optical depth of the layer equal unity (see eq.~[\ref{eq:tau1}] before), in effect rendering $\mathcal{A}_{\nu}^{\rm th} = \mathcal{A}_{\nu}$ a function of $\eta$ only.     

On the other hand, when the ejecta is fully ionized, then any photons that pass through the layer escape to infinity and hence are lost from the system, i.e.~they are not deposited in the thermal bath.  In this case $\mathcal{A}_{\nu}^{\rm th} \lesssim \mathcal{A}$ and each are calculated separately as a function of both $\eta$ and for a layer thickness with the same optical depth $\tau_{\rm es}^{\rm ej}$ as that of the total ejecta (eq.~[\ref{eq:tau_ej}]). The luminosity of escaping non-thermal radiation (third term in eq.~[\ref{eq:photon_evo}]) is thus given by
\be
L_{\rm nth,\nu} = (\mathcal{A}_{\nu}-\mathcal{A}_{\nu}^{\rm th})\frac{E_{\rm nth,\nu}}{t_{\rm d}^{\rm n}}
\label{eq:Lnth}
\ee
In calculating the X-ray luminosity $L_{X}$ we integrate $L_{\rm nth,\nu}$ over a particular band of interest, e.g., $\sim 1-10$ keV.

Equation (\ref{eq:photon_evo}) neglects any change in the non-thermal radiation of the nebula due to Compton scattering by the background nebular pairs.  This is a good approximation if the Compton $y$-parameter, which is given by
\be
y = \frac{4 kT_{\pm}\tau_{\rm es}^{\rm n}}{m_e c^{2}} \approx 8\times 10^{-3}\left(\frac{kT_{\pm}}{\rm keV}\right)\tau_{\rm es}^{\rm n},
\label{eq:y}
\ee
obeys $y \lesssim \beta$, since this condition is equivalent to the timescale for photon-pair equilibration being shorter than the evolution timescale of the nebula (e.g.,~\citealt{Metzger+13}).  Here $T_{\pm}$ is the Compton temperature of the pairs, normalized to a typical value $\sim 1$ keV.  Equation (\ref{eq:y}) shows that $y \lesssim \beta \sim 1$ is violated only at very early times $\ll t_{\rm d,0}^{\rm ej}$, well before the peak of the optical and X-ray light curves.  Nebular pairs cannot therefore thermalize the nebular radiation field at most times of interest.  

\subsection{Ionization State of Ejecta}
\label{sec:ionization}
The depth of the ionized layer $\Delta^{i}$ for each ionization state of relevance is set by the location at which the optical depth of a photon of frequency $\nu \gtrsim \nu_{\rm thr}^{i}$ to absorption reaches unity, i.e.,
\be
1 = \int_{0}^{\Delta^{i}}\rho_{\rm ej}\kappa_{\rm bf,\nu}^{i}\left[1 +\rho_{\rm ej}\kappa_{\rm es}s \right]ds \approx \tau_{\rm abs}^{i}(1 + \tau_{\rm es}^{i}),
\label{eq:tau1}
\ee
where $\nu_{\rm thr}^{i}$ is the threshold frequency for the $i$th state, $s$ is the depth through the ionized layer, $\rho_{\rm ej} = M_{\rm ej}/2m_p V$ is the density of the shocked gas, and 
\be
\tau_{\rm abs}^{i} \equiv \kappa_{\rm bf,\nu}^{i}\rho_{\rm ej}\Delta^{i}
\ee
is the optical depth through the layer to absorption, where 
\be
\kappa_{\rm bf,\nu}^{i} \simeq \frac{f_{\rm n}^{i}}{56 m_p}\sigma_{\rm bf,\nu}^{i}
\label{eq:kappa}
\ee
is the absorption opacity and $\sigma_{\rm bf,\nu}^{i}$ is the bound-free cross section from Verner et al.~(1996).  The factor $1 + \tau_{\rm es}^{i}$ in equation (\ref{eq:tau1}) accounts for the additional path-length traversed by the photon due to electron scattering,
where
\be
\tau_{\rm es}^{i} = \rho_{\rm ej}\kappa_{\rm es}\Delta^{i}
\ee
is the electron scattering optical depth through the ionized layer.  The neutral fraction $f_n^{i}$ is determined by the balance between ionization and recombination
\be
f_n^{i} = \left(1 + \frac{4\pi}{\alpha_{\rm rec}^{i} n_e}\int\frac{J_{\nu}}{h\nu}\sigma^{i}_{\rm bf}(\nu)d\nu\right)^{-1},
\label{eq:fn}
\ee
where $J_{\nu} = c E_{\rm nth, \nu}/4\pi V$ is the mean intensity inside the nebula; $n_e \approx \rho_{\rm sh}/2m_p$ is the electron density of the ejecta.  Here $\alpha_{\rm rec}^{i}$ is the radiative recombination rate coefficient (e.g.,~\citealt{Nahar&Pradhan94}; \citealt{Nahar06}; \citealt{Nahar+09}; \citealt{Nahar+11}), which depends on the temperature of the electrons $T_e^{i}$ in the ionized layer.  The electron temperature is set by Compton equilibrium, which also depends on the spectrum of photons present in the ionizing layer as described in \citet{Metzger+13}.

Since some species are more easily ionized than others, only a limited set of ionization states of iron are relevant in setting the ionization depth across a given range of photon energies.  These most important ions are determined by first calculating $\Delta^{i}$ for {\it all} ionization states of iron, assuming that photons at $\nu \sim \nu_{\rm thr}^{i}$ propagate through the ejecta unattenuated, except due to absorption by species $i$.  Then, starting at low frequencies and moving to progressively higher frequencies, we `eliminate' those species with a larger value of $\Delta^{i}$ than the penetration depth allowed by the previous element with the lower ionization frequency.  Applying this procedure results in the following set of {\it relevant} ionization states: Fe$^{1+}$ (8$-$31 eV), Fe$^{3+}$ (31$-$99 eV), Fe$^{6+}$ (99-1360 eV), Fe$^{18+}$ (1.36-8.8 keV), Fe$^{25+}$ ($> 8.8$ keV).

Although our simple-minded procedure is ultimately no substitute for a time-dependent photo-ionization calculation, it nevertheless provides a basic estimate of what conditions are required for UV/X-ray photons of a given energy to directly escape the nebula.  One of the biggest limitations of our model in the present context is that it assumes that ionization balance is achieved instantaneously as the conditions in the ejecta change.  This is a good approximation at times $t \gtrsim t_{\rm d,0}^{\rm ej}$ (eq.~[\ref{eq:td}]) since this it the time after which the diffusion time is shorter than the evolution timescale.  Many of our calculations appear to indicate that full ionization occurs earlier than this time, which is not possible since the ionizing photons have not yet had time to diffuse to the surface via Thomson scattering.  In these cases we artificially delay full ionization until $t \sim t_{\rm d,0}^{\rm ej}$, after which time we are confident that the ionizing photon flux has had sufficient time to reach the ejecta surface.

\section{Numerical Results}
\label{sec:results}

In all cases we assume that the pulsar has an initial rotation period $P \simeq 2$ ms for reasons already discussed.  The magnetic field of the merger remnant, being much less certain, is varied across the range of values $B_{\rm d} \sim 10^{14}-10^{16}\,{\rm G}$.  We consider ejecta masses $\sim 0.01-0.1M_{\odot}$ within the expect range (\citealt{Hotokezaka+13};  \citealt{Fernandez&Metzger13}) and always assume an initial velocity $\beta_0 = 0.1$.  Our results are insensitive to this assumption because the ejecta kinetic energy is dominated at times of interest by the fixed total rotational energy deposited by the pulsar at much earlier times $\lesssim t_{\rm sd}$.

\begin{figure}
\includegraphics[width=0.5\textwidth]{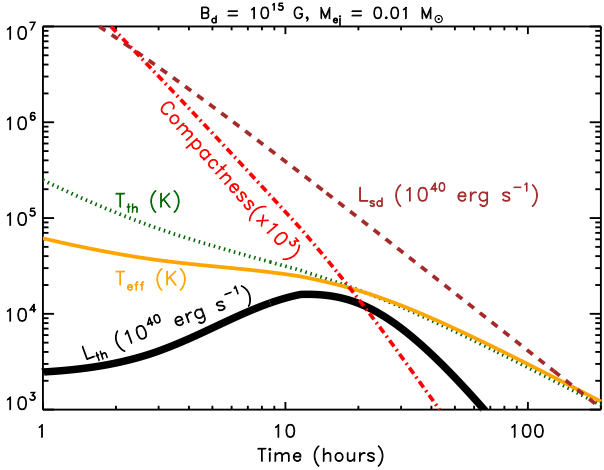}
\caption{Various quantities as a function of time since the merger, calculated for a pulsar with dipole field $B_d = 10^{15}$ G and initial rotation period $P = 2$ ms.  We  assume ejecta of mass $M_{\rm ej} = 10^{-2}M_{\odot}$ and initial velocity $v_{\rm ej,0} = 0.1$ c. Quantities shown include the spin-down luminosity of the pulsar $L_{\rm sd}$ (dashed brown; eq.~[\ref{eq:Lsd}]), the bolometric luminosity of the thermal emission $L_{\rm th}$ (solid black; eq. [\ref{eq:L_SN}]), effective temperature of the thermal radiation $T_{\rm eff} \equiv (L_{\rm th}/4\pi \sigma R_{\rm ej}^{2})^{1/4}$, temperature of the thermal bath in the nebula $T_{\rm th}$ (dotted green), and compactness of the nebula $\ell$ (red dot-dashed; eq. [\ref{eq:compactness}]).
} 
\label{fig:evo15}
\end{figure}

Figure \ref{fig:evo15} summarizes the basic characteristic features of one of our calculations, in this case for a magnetar with $B_{\rm d} = 10^{15}$ G and ejecta mass $M_{\rm ej} = 10^{-2} M_{\odot}$.  The thermal luminosity ({\it black line}) reaches its peak value $L_{\rm th, peak} \sim 2\times 10^{44}\,{\rm erg\, s^{-1}}$ on a timescale $\sim 15\,{\rm hours}$, with an effective temperature $T_{\rm eff} \sim 2\times 10^{4}\,{\rm K}$, corresponding to a spectral peak in optical/UV.  The peak timescale is set by the characteristic diffusion timescale of the ejecta $t_{\rm d,0}^{\rm ej}$ for $\beta \sim 1$ (eq.~[\ref{eq:td}]).  The nebula compactness remains $\ell \gg 1$ until times $t \gtrsim 40\,{\rm hours}$, justifying our assumption of a saturated pair cascade and the resulting flat X-ray spectrum.  

\begin{figure}
\includegraphics[width=0.5\textwidth]{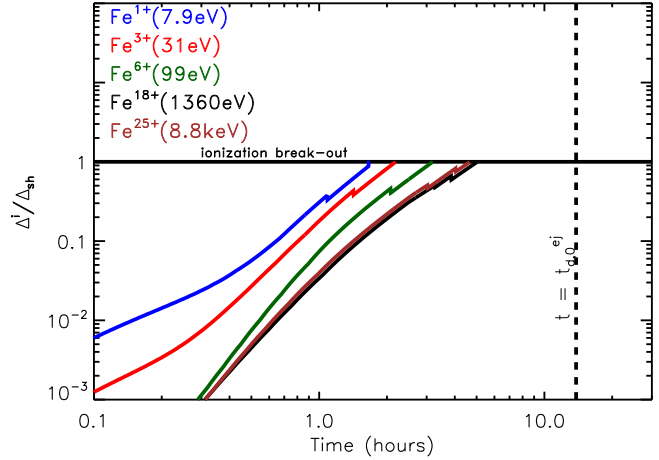}
\caption{Thickness of the ionized layer $\Delta^{i}$ relative to the total width of the ejecta $\Delta_{\rm sh}$ ($\S\ref{sec:ionization}$) for those ions that dominate the ionization front in a given photon energy range, for the same calculation shown in Figure \ref{fig:evo15}. Different ionization states of iron are denoted by different colors as indicated. Ionization break-out ($\Delta^{i}/\Delta_{\rm sh} = 1$) occurs first at UV energies and later at soft X-ray energies.  Since physically the ionization front cannot propagate through the ejecta on a timescale shorter than $t_{\rm d,0}^{\rm ej}$ ({\it vertical dashed line}), our calculation actually underestimates the timescale for the onset of the UV/X-ray emission from the nebula, which will instead begin at times $t \gtrsim t_{\rm d,0}^{\rm ej}$.  As described in the text, in such cases we artificially delay full ionization until $t \sim t_{\rm d,0}^{\rm ej}$, after which time we are confident that the ionizing photon flux has had sufficient time to reach the ejecta surface.}
\label{fig:ion15}
\end{figure}

The peak thermal luminosity is considerable: {\it over two orders of magnitude higher than that of a normal kilonova powered by radioactive decay} (for which $L_{\rm th, peak} \lesssim 10^{42}\,{\rm erg\, s^{-1}}$).  This emission is nevertheless much dimmer than the pulsar luminosity at the time of peak emission, in contradiction to the prediction of simple theoretical models that deposit the pulsar energy directly into the thermal bath of the nebula (e.g.,~\citealt{Kasen&Bildsten10} in the context of supernovae and \citealt{Yu+13} in the context of NS mergers).  This low radiative efficiency is the combined result of two main effects, (1) the low efficiency with which the ejecta absorbs nebular X-rays (albedo $> 0$) as the ejecta becomes fully ionized, as discussed in \S\ref{sec:ionization}, and (2) the optical depth of pairs in the nebula, which we discuss next.

A key parameter in quantifying the impact of the nebular pairs is the ratio of the characteristic diffusion timescale through the nebula (eq.~[\ref{eq:tdiffn}]) and that through the ejecta (eq.~[\ref{eq:td}]) (Appendix \ref{sec:appendixLC}):
\be
\chi \equiv \frac{t_{\rm d,0}^{\rm n}}{t_{\rm d,0}^{\rm ej}} \simeq 2.7 B_{15}^{-2/3}M_{-2}^{-1/2}\beta^{5/6} \approx 2.7 B_{15}^{-2/3}M_{-2}^{-11/12},
\ee
where in the second equality we have used the approximate relationship between ejecta velocity and mass, $\beta \simeq 1.0M^{-1/2}_{-2}$.  When $\chi > 1$, then X-rays require longer to diffuse across the nebula than it takes for photons to diffuse out of the ejecta around the time of peak emission.  In this case, X-ray photons produced by the pulsar wind lose a portion of their energy to PdV expansion before they can cross the nebula to have their energy absorbed by the ejecta walls (see Figures \ref{fig:schematicLC} and \ref{fig:LCs}).  In Appendix \ref{sec:appendixLC}, we show that this effect alone suppresses the peak luminosity by a factor $\chi^{-3/2}$ even if there was 100$\%$ efficient absorption by the ejecta walls (albedo = 0).

Figure \ref{fig:ion15} shows the evolution of the thickness of ionization fronts of various ionization states of iron $\Delta^{i}$ as a fraction of the thickness of the ejecta $\Delta_{\rm sh}$, for the same model shown in Figure \ref{fig:evo15}.  The front comprised of Fe$^{1+}$ reaches the ejecta surface first on a timescale $\gtrsim 1$ hour, after which time UV photons (8-31 eV) are free to escape the nebula through the ejecta without being absorbed.  Fronts associated with sequentially higher ionization states (Fe$^{3+}$-Fe$^{25+}$) reach the ejecta surface over the next few hours, after which time X-rays are also free to escape the ejecta.  As described previously, complete ionization at times $\ll t_{\rm d,0}^{\rm ej}$ cannot actually occur because ionizing photons propagate through the ejecta at a rate that is ultimately limited by the electron scattering diffusion timescale, which is much longer than the evolution time when $t \ll t_{\rm d,0}^{\rm ej}$.  Our model, which assumes a steady-state ionization structure, does not capture this effect.  The ionization break-out that we find occurs at times $\lesssim t_{\rm d,0}^{\rm ej}$ should instead be interpreted as indicating that complete ionization, and hence the onset of non-thermal radiation, will occur on a timescale $t_{\rm d,0}^{\rm ej} \sim$ 14 hours.  The X-ray emission peaks around this time at a luminosity $L_{\rm X, peak} \sim 3\times 10^{44}\,{\rm erg\, s^{-1}}$.

\begin{figure}
\includegraphics[width=0.5\textwidth]{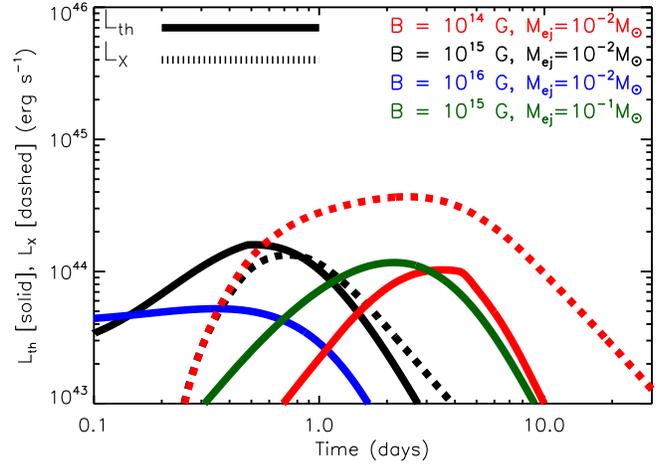}
\caption{Thermal optical/UV ({\it solid}) and non-thermal X-ray ({\it dashed}) luminosity as a function of time, for different models as labeled with different colors.  X-ray emission is only observable for those systems for which the nebular X-ray luminosity is sufficient to ionize the ejecta.  This condition is satisfied by the $B_{\rm d} = 10^{14}$ G and $B_{\rm d} = 10^{15}$ G models with $M_{\rm ej} = 10^{-2}M_{\odot}$ but not by the other two models. }
\label{fig:LCrange}
\end{figure}

To better explore the range of possible signatures, we perform a range of calculations similar to those shown in Figures \ref{fig:evo15} and \ref{fig:ion15} by varying the magnetic field strength $B_{\rm d}$ and ejecta mass $M_{\rm ej}$ across the range of physical values.  Figure \ref{fig:LCrange} shows our results for the thermal optical/UV and non-thermal X-ray light curves ($1-10\,{\rm keV}$).  Across this relatively wide parameter space, the thermal luminosity peaks at $\sim 10^{43}-10^{44}$ erg s$^{-1}$ on a timescale of several hours to days.   In cases of a relatively weak magnetic field $B = 10^{14}-10^{15}$ G and relatively low ejecta mass $M_{\rm ej} = 10^{-2}M_{\odot}$, the ejecta is fully ionized by nebular photons, allowing non-thermal emission to be observed at times $t \gtrsim t_{\rm d,0}^{\rm ej}$.  When present, the resulting X-ray luminous is comparable, or moderately greater than, the thermal luminosity.   For a particularly strong magnetic field of $B = 10^{16}$ G or a high ejecta mass $M_{\rm ej} = 0.1 M_{\odot}$, the ejecta is not fully ionized on timescales of interest.  Since in this case the non-thermal X-rays remain trapped behind the neutral ejecta, no X-ray emission observable.  For the reader interested in additional details, analytic estimates of the optical and X-ray light curves are presented under idealized assumptions in Appendix \ref{sec:appendixLC}, while an analytic estimate of the conditions required for ejecta ionization is given in Appendix \ref{sec:ion}.

\begin{figure}
\includegraphics[width=0.5\textwidth]{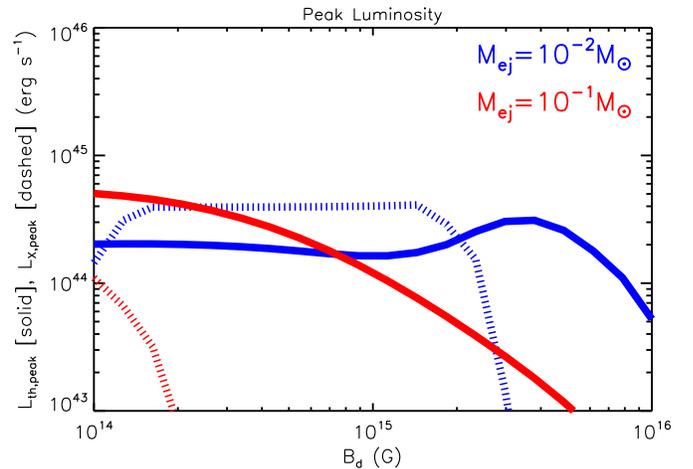}
\caption{Peak thermal ({\it solid}) and X-ray (0.1-1 keV; {\it dashed}) luminosity as a function of the dipole field strength $B_{\rm d}$, calculated for two values of the ejecta mass $M_{\rm ej} = 10^{-2}M_{\odot}$ ({\it blue}) and $M_{\rm ej} = 10^{-1}M_{\odot}$ ({\it red}).}

\label{fig:models_lum}
\end{figure}

Figure \ref{fig:models_lum} shows the peak thermal and X-ray luminosities, calculated from a continuous series of models with $10^{14}\,{\rm G}\le B_{\rm d}\le 10^{16}\,{\rm G}$ and for two values of the ejecta mass, $M_{\rm ej} = 10^{-2}$ ({\it blue}) and $10^{-1}M_{\odot}$ ({\it red}).  The peak optical luminosity is smaller for larger values of $B_{\rm d}$ due primarily to the lower spin-down luminosity at times $t \gg t_{\rm sd}$, resulting in most of the rotational energy of the pulsar being lost to PdV work.  The peak X-ray luminosity also decreases for larger $B_{\rm d}$ (eq.~[\ref{eq:LpeakX}]) and in fact drops to zero above a critical magnetic field strength, which is approximately $B_{\rm d} \sim 10^{14}(2\times 10^{15})$ G for the $M_{\rm ej} = 10^{-1}(10^{-2})M_{\odot}$ cases, respectively.  As discussed above, this X-ray shut-off occurs once the nebular X-ray luminosity is insufficient to ionize the ejecta on timescales of interest (near the optical peak).  In Appendix \ref{sec:ion} we derive a simple analytic estimate of the conditions required for complete ionization of the ejecta, from which we show that the time of ionization is a strongly increasing function of $B_{\rm d}$ and $M_{\rm ej}$ (eq.~[\ref{eq:tbo2}]).  

\section{Discussion}
\label{sec:discussion}

\subsection{Implications for X-ray emission following short GRBs}

Many short GRBs are accompanied by temporally extended X-ray emission in excess of that expected from the normal non-thermal afterglow produced by the GRB jet.  This includes bright X-ray flaring on timescales $\sim$ 10-100 seconds after the initial spike (e.g.,~\citealt{Norris&Bonnell06}; \citealt{Bostanci+13}), which could be powered by jetted emission from a collimated magnetar outflow (\citealt{Metzger+08}; \citealt{Bucciantini+12}; \citealt{Gompertz+13}), a process qualitatively similar to jetted emission responsible for normal prompt GRB emission.  Other short GRBs are accompanied by X-ray emission with plateau-like light curve (e.g.,~\citealt{Rowlinson+13}).  This emission could also be powered by magnetar spin-down energy directly (e.g.,~\citealt{Metzger+11}; \citealt{Rowlinson+13}), or indirectly via energy injection to the afterglow shock (e.g.,~\citealt{DallOsso+11}).

Both direct emission from the magnetar jet and from a magnetar-rejuvenated afterglow require that the magnetar outflow is effectively collimated into a bipolar jet even at late times when the spin-down power is low (which, as we have argued, may be unlikely; eq.~[\ref{eq:jbo}]).  In this paper we have instead focused on emission produced by the diffusion of radiation from the magnetar-powered nebula outwards through the expanding ejecta.  For non-thermal X-ray emission to be observed in this scenario, the nebular radiation must be sufficient to completely ionize the ejecta and one must wait until a time $\gtrsim t_{\rm d,0}^{\rm ej} \sim$ several hours to day.  

If accretion onto the magnetar, or an early collimated jet from the magnetar itself, can power a short GRB, then optical/X-ray emission from the subsequent magnetar nebula phase could in principle be detected on several hours to days following the GRB as we have investigated.  To be detectable, such emission from the magnetar must exceed that of the GRB afterglow.  Approximately half of short GRBs are detected in the optical on timescales $\sim$ hours after the burst, with typical luminosities/upper limits $L_{\rm opt} \sim 10^{42}-10^{45}$ erg s$^{-1}$ (see \citealt{Metzger&Berger12} Fig.~3 for a compilation).  Since these are comparable to the peak thermal luminosities expected $L_{\rm th,peak} \sim 10^{43}-10^{45}$ erg s$^{-1}$ (Fig.~\ref{fig:LCrange}, \ref{fig:models_lum}), magnetar emission should be detectable above the afterglow emission in at least in a fraction of systems. {\it This wold provide a clear signal that a magnetar indeed survived following the NS merger.}  We next discuss two specific examples that appear to show many of the features we expect for such events.

\subsubsection{GRB 080503}

GRB 080503 was a short GRB with bright extended X-ray flaring lasting $\sim 100$ seconds after the burst.  This event showed an unusual rise in its afterglow light curve at $t \sim 1$ day, before fading over the next several days (\citealt{Perley+09}).  The observed evolution of this optical emission was broadly consistent with that expected from a kilonova powered by radioactive decay (assuming the radioactive matter possessed an iron-like opacity to give the right timescale).  However, although the event was well localized on the sky, no obvious host galaxy was detected coincident with the burst, despite the fact that a relatively low redshift ($z \lesssim 0.2$) would be required to produce the expected peak brightness of a kilonova. For this reason, a kilonova interpretation was disfavored for explaining the rebrightening.

In principle the optical rebrightening following GRB 080503 could be powered by energy injection from a central magnetar, as in the model proposed here.  Since the optical peak luminosity is likely to be much higher in this case than in a standard kilonova, this would place the burst at a higher redshift more consistent with the absence of a host galaxy (although this is hardly a ironclad argument, as some GRBs can occur far outside their hosts; e.g.,~\citealt{Fong&Berger13}).   Also, although the X-ray afterglow of the GRB itself faded rapidly after the burst, X-ray emission was detected near the optical peak with a luminosity comparable to the optical emission. Again this would not be expected for a typical kilonova, but it is consistent with the emission we expect from the the magnetar wind nebula.  Given its bright extended X-ray emission, and its later optical/X-ray emission brightening, GRB 080503 is a good candidate for a magnetar-powered transients as described here.  \citet{Gao+13} present a similar interpretation for this event, although the details of our model differ significantly as we described earlier.

\subsubsection{GRB 130603B}

The short GRB 130603B recently received attention for showing an infrared excess in its late emission, consistent with an $r$-process powered kilonova (\citealt{Berger+13}; \citealt{Tanvir+13}).  Early fits to the optical and radio data showed evidence for a jet break at $t \sim 12$ hours \citep{Fong+13}, making this one of only a handful of bursts with a measured jet opening angle.  However, the X-ray light curve showed an excess of approximately $L_X \simeq 4\times 10^{43}(t/{\rm day})^{\alpha}$ erg s$^{-1}$ where $\alpha \simeq -1.88$ at times $t \gtrsim$ few hours (\citealt{Fong+13}).   Comparing this to our models, such emission is consistent with a magnetar with $B_{\rm d} \simeq 3\times 10^{15}$ G and $M_{\rm ej} = 10^{-2}M_{\odot}$ (Fig.~\ref{fig:evo15}).  Emission from the magnetar nebula cannot explain the X-ray emission at early times ($t \lesssim 10$ hours), since the ejecta has not yet been ionized and because the X-rays have not yet had time to diffuse through the ejecta.  The early X-ray emission could be associated with the normal GRB afterglow, with the magnetar emission becoming dominant once the afterglow decays following the jet break.  The optical/IR emission from the magnetar remnant is predicted to be well below the observed optical emission, suggesting that the optical emission likely originated from the afterglow itself (\citealt{Cucchiara+13}) and is not associated with the thermal emission of the model we have investigated.

\subsection{Late-time radio emission}

Another way to test for the presence of a long-lived magnetar is via late radio emission. This may arise either from the nebula itself on the timescale of $\sim\,{\rm months}$ \citep{Piro&Kulkarni13} or from the interaction of the relativistic ejecta with the surrounding circumburst medium on the timescale of $\sim\,{\rm years}$ \citep{Metzger&Bower13}.  \citet{Metzger&Bower13} conducted a search for late radio emission following seven short GRBs on timescales $\gtrsim 1-3\,{\rm  years}$ after the bursts with the Very Large Array.  They detected no sources, which they used to rule out the presence of a millisecond magnetar remnant in two short GRBs (one, GRB 050724, also shows temporally extended X-ray emission), although the other systems were less constrained due to the uncertain density of the circumburst medium.  Unfortunately, the observations were not conducted soon enough after the GRBs to constrain the nebular emission models, which would have required $\sim{\rm month}$ timescale followup. Future more sensitive observations with the recently upgraded VLA (especially in cases of known high circumburst densities for interaction models) could be used to place significantly tighter constraints on the presence of long-lived magnetar remnants in short GRBs.

\subsection{Implications for the radioactively-powered red kilonova emission from the tidal tail}
\label{sec:lanthanides}

We have assumed ejecta with an Fe-like composition, since our focus here was on the more spherical outflows from the accretion disk.  However, a comparable amount of mass is ejected dynamically from the merger in equatorially-concentrated tidal tails, and this material is instead composed of heavy $r$-process elements.  \citet{Kasen+13} (cf.~\citealt{Barnes&Kasen13}; \citealt{Tanaka&Hotokezaka13}) recently showed that the Lanthanide elements likely produce a much higher opacity than Fe-like nuclei, due to the large number of bound-bound line transitions associated with their valence f-shell electrons.   

We have shown that X-rays within the nebula are sufficient to {\it completely} ionize iron-like nuclei at times $\gtrsim$ several hours to days, depending on the ejecta mass and the magnetic field of the pulsar.  This suggests it is possible that X-rays from the magnetar nebula may in some cases be sufficient to ionize the valence electrons of the Lanthanide elements.  Since these electrons are the source of the high opacity, this would substantially change the predicted light curves and spectra of the resulting kilonova.  

As already discussed, GRB 130603B was accompanied by red emission at $t \sim 9$ days indicative of an $r$-process powered kilonova with an inferred ejecta mass $M_{\rm ej} \sim 3\times 10^{-2}M_{\odot}$ (\citealt{Tanvir+13}; \citealt{Berger+13}).  Above we showed that the excess X-ray emission of this event was well-fit by a magnetar, assuming a dipole field $B_{\rm d} \sim 3\times 10^{15}$ G.  If we substitute these numbers into our analytic expression for the ionization time derived in the Appendix \ref{sec:ion} (eq.~[\ref{eq:tbo2}]), we calculate an ionization time of $\sim 140$ days, much larger than the time of the observed IR excess.  This calculation was, however, made for hydrogen-like iron, not the singly or doubly ionized Lanthanides of interest for the opacity.  A more detailed calculation, requiring the bound-free cross section and recombination coefficients of the Lanthanides and accounting for the more complex geometry of the tidal tails, is necessary to determine whether the X-rays observed in GRB were indeed sufficient to ionize the Lanthanides.  This issue is of critical importance as it bears on whether the IR emission from this event was indeed a kilonova.  

\subsection{Implications for gravitational wave follow-up}

The transients we predict are important as potential electromagnetic counterparts to the gravitational waves produced by the inspiral chirp of coalescing NSs.  Although a rather stiff equation of state is required to support an indefinitely stable NS remnant, this effect is offset by the fact that population synthesis predicts a large number of low mass mergers which could leave such remnants (e.g.,~\citealt{Belczynski+08}). On one hand, our electromagnetic signature has the same important use of any other electromagnetic counterpart in assisting with localization and identification of the environment where the merger occurred \citep{Metzger&Berger12,Nissanke+13,Kasliwal&Nissanke13}. This allows one to maximize the astrophysics that can be learned from any given event. On the other hand, the detection or non-detection of the signatures we predict would be a critical test of NS equations of state \citep{Hebeler+13}. The gravitational wave detections provide a measurement of the NS masses, and once the second-generation gravitational-wave interferometers currently under construction \citep{LIGO,VIRGO,KAGRA} begin regularly detecting NS mergers, the occurrence or non-occurrence of a NS remnant would indicate what the maximum mass of a NS must be.

\subsection{Blind detection with optical transient surveys}

The optical peak luminosities $\sim 10^{44}$ erg s$^{-1}$ that we predict from magnetar merger remnants are $\gtrsim 10$ times greater than those of normal supernovae, making them detectable to cosmological distances.  Even though the rates of binary NS mergers are a factor $\gtrsim 10-100$ times lower than that of supernovae (and the fraction of mergers producing stable remnants may be small) such events are promising targets for discovery with wide-field optical transient surveys.  Given the predicted short duration of the emission $\sim$ several hours to days compared to supernovae, surveys with particularly rapid cadences (ideally $\lesssim$ hours-day), such as the Zwicky Transient Facility (\citealt{Law+09}) or the ASAS-SN (\citealt{Shappee+13}) are the most likely to be successful.  Note that \citet{Wu+13} propose a model qualitatively similar to that described here to explain the rapidly fading non-thermal transient PTF11agg (\citealt{Cenko+13}).  We encourage future high cadence deep surveys to detect or constrain the rates of these rare, luminous, and short-lived events.

\subsection{White dwarf accretion-induced collapse}

A model physically similar to that we have developed for NS merger remnants may also apply to a different astrophysical event: the accretion-induced collapse (AIC) of a white dwarf.  AIC occurs once a white dwarf of O-Ne composition accretes up to near the Chandrasekhar mass. due to the loss of degeneracy support resulting from electron capture (e.g.~\citealt{Nomoto&Kondo91}).  The accreted mass may originate from steady Roche lobe overflow from a non-degenerate companion star, or it may originate from the merger of a double white dwarf binary with total mass above the Chandraskehar limit (e.g.~\citealt{Tauris+13}).  Because a WD accretes angular momentum as well as mass, it is typically rotating rapidly prior to AIC.  As a result, the outer layers of the collapsing WD may form an accretion disk around the newly-formed NS (\citealt{Dessart+06}; \citealt{Abdikamalov+10}), the properties of which depend sensitively on the rotation rate (and the degree of differential rotation; \citealt{Piro08}) in the pre-collapse white dwarf.  

The configuration following AIC, namely a rapidly rotating neutron star surrounded by a small quantity of outflowing material, thus qualitatively resembles that following a NS merger leaving a long-lived NS remnant.  Assuming that strong magnetic fields are also generated in these events (\citealt{Usov92}), AIC could produce optical/X-ray emission in a manner very similar to that described here.  As in the case of NS mergers, such emission would likely outshine any radioactively-powered emission from these events (\citealt{Metzger+09b}; \citealt{Darbha+10}).  A potentially important difference between AIC and NS mergers is the initial rotation rate of the NS, which in AIC could be substantially less than the centrifugal break-up limit if the WD is rotating relatively slowly at the point of collapse (\citealt{Thompson&Duncan95}).  By contrast, in the merger case, the large angular momentum of the initially binary virtually ensures that the initial spin period be less than a few milliseconds.    

\section{Conclusions}
\label{sec:conclusions}

We have explored the optical and X-ray signatures of a long-lived magnetar remnant creating following a NS binary merger.  Our conclusions are summarized as follows.

\begin{itemize}
\item{Energy injection from a stable millisecond magnetar formed from a binary NS merger can produce a bright thermal UV/optical and X-ray transient, powered by the diffusion of radiation outwards from a millisecond magnetar nebula.  The emission is present even if the magnetar outflow is stifled behind the merger ejecta, as appears likely at late times as the pulsar luminosity decreases (eq.~[\ref{eq:jbo}] and surrounding discussion).}
\item{We have applied the model of \citet{Metzger+13} to the evolution of millisecond magnetar wind nebulae produced following mergers, taking into account the evolution of the thermal and non-thermal radiation, and their coupling via absorption on the ejecta walls.  Stages of the model are summarized in Figure \ref{fig:schematic}.}
\item{In most cases the X-ray luminosity of the nebula is sufficient to ionize the ejecta, such that non-thermal emission can escape the nebula on the photon diffusion timescale set by Thomson scattering (Fig.~\ref{fig:ion15}; Appendix \ref{sec:ion}).}
\item{Thermal emission from the nebula peaks on a timescale of several hours to days at a luminosity $L_{\rm th,peak} \sim 10^{43}-10^{45}\,{\rm erg\,s^{-1}}$ with effective temperatures corresponding to UV/optical frequencies.  This is a factor $\gtrsim 100$ times brighter than the emission expected from a radioactively powered kilonova. }
\item{The optical emission is substantially less (by a factor of $\sim100$) than that predicted by simpler models which assume (without justification) a fixed fraction of the pulsar luminosity thermalizes (e.g.,~\citealt{Yu+13}). The main reasons for this suppression are (1) the non-zero albedo of the ejecta walls, and (2) the high optical depth of pairs through the nebula. Appendix \ref{sec:appendixLC} and Figure \ref{fig:LCs} summarize these effects.}
\item{The X-ray luminosity peaks at $L_{\rm peak,X} \sim 10^{43}-10^{45}\,{\rm erg\, s^{-1}}$, similar to the thermal luminosity, and declines at late times proportional to the pulsar luminosity $\propto t^{-2}$.}
\item{Emission from a long-lived magnetar remnant could in principle be observed following short GRBs, depending on whether it can be detected above the optical/X-ray afterglow. This would be important evidence that a massive NS was created in the merger rather than a BH. The optical rebrightening and X-ray emission from GRB 080503, and the X-ray excess following GRB 130603B, are consistent with being powered by a magnetar.}
\end{itemize}

\section*{Acknowledgments}
We are grateful to Todd Thompson for many helpful discussions and for detailed comments on the manuscript.  We thank Andrei Beloborodov, Indrek Vurm, and Romain Hascoet for their help in developing the model for millisecond PWN adapted for use in this paper.  We thank John Beacom, He Gao, Eliot Quataert, and Bing Zhang for helpful discussions.  ALP is supported through NSF grants AST-1205732, PHY-1068881, PHY-1151197, and the Sherman Fairchild Foundation.

\appendix

\section{Analytic Estimate of Light Curve}
\label{sec:appendixLC}


The diffusion time through the nebula and ejecta evolve, respectively, with time as
\begin{eqnarray}
t_{\rm d}^{\rm ej} = \frac{R}{c}(\tau_{\rm es}^{\rm ej}+1) = \left\{
\begin{array}{lr}
t_{\rm d,0}^{\rm ej}\left(\frac{t}{t_{\rm d,0}^{\rm ej}}\right)^{-1} , 
& t \lesssim t_{\rm thin}^{\rm ej} \\
\beta t
,& t \gtrsim t_{\rm thin}^{\rm ej} \\
\end{array}
\right.
\label{eq:tejcases}
\end{eqnarray}
\begin{eqnarray}
t_{\rm d}^{\rm n} = \frac{R}{c}(\tau_{\rm es}^{\rm n}+1) =  \left\{
\begin{array}{lr}
t_{\rm d,0}^{\rm n}\left(\frac{t}{t_{\rm d,0}^{\rm n}}\right)^{-1/2} , 
&  t \lesssim t_{\rm thin}^{\rm n} \\
\beta t
,& t \gg t_{\rm thin}^{\rm n}, \\
\end{array}
\right.
\label{eq:tdncases}
\end{eqnarray}
where $t_{\rm d,0}^{\rm  ej}$ (eq.~[\ref{eq:td}]) and $t_{\rm d,0}^{\rm n}$ (eq.~[\ref{eq:tdn}]) are the characteristic timescales at which the diffusion time equals the expansion time.  The different temporal evolution of $t_{\rm d}^{\rm ej} \propto t^{-1}$ and $t_{\rm d}^{\rm n} \propto t^{-1/2}$ results because although the number of electrons in the ejecta is fixed, the number of e$^{+/-}$ pairs in the nebula changes due to the evolving balance between injection and annihilation (eq.~[\ref{eq:Npmevo}]). 

The ratio of the characteristic diffusion times forms a critical parameter
\be
\chi \equiv \frac{t_{\rm d,0}^{\rm n}}{t_{\rm d,0}^{\rm ej}} \simeq 2.7 B_{15}^{-2/3}M_{-2}^{-1/2}\beta^{5/6} \simeq 2.7 B_{15}^{-2/3}M_{-2}^{-11/12}.
\label{eq:chi}
\ee
The ejecta velocity is sufficiently close to the speed of light in most cases of interest that ejecta becomes optically thin at a time $t_{\rm thin}^{\rm ej} = t_{\rm d,0}^{\rm ej}\beta^{-1/2}$ similar to $t_{\rm d,0}^{\rm ej}$.  Thus, depending on the value of $\chi$, we have two possible orderings of the times: (1) $t_{\rm d,0}^{\rm n} \lesssim t_{\rm d,0}^{\rm ej} \sim t_{\rm thin}^{\rm ej}$ when $\chi < 1$, and (2) $t_{\rm d,0}^{\rm ej} \sim t_{\rm thin}^{\rm ej} \lesssim t_{\rm d,0}^{\rm n}$ when $\chi > 1$.  Physically, the cases $\chi < 1$ and $\chi > 1$ correspond to whether the nebula or ejecta becomes free-streaming first, respectively.

Now consider a frequency-integrated version of the evolution equation (eq.~[\ref{eq:photon_evo}]) for the non-thermal photons in the nebula
\be
\frac{dE_{\rm nth}}{dt} =  -\frac{E_{\rm nth}}{R}\frac{dR}{dt} + L_{\rm sd} - (1-A)\frac{E_{\rm nth}}{t_{\rm d}^{\rm n}} ,
\label{eq:photon_evo2}
\ee
where $E_{\rm nth} = \int E_{\rm nth,\nu}d\nu$ is the total non-thermal energy and $A = \int (\mathcal{A}_{\nu}-\mathcal{A}_{\nu}^{\rm th})d\nu/\int d\nu$ is a frequency-averaged albedo (probability of reflection instead of absorption or escape).  For the time being we neglect energy loss to radiation $L_{\rm nth}$ in equation (\ref{eq:photon_evo2}).  

Although the pulsar luminosity $L_{\rm sd}$ represents the only source in (\ref{eq:photon_evo2}), the dominant loss term changes with time.   At early times $t \ll t_{\rm d,0}^{\rm n}$ the dominant loss term is PdV work $\sim E_{\rm nth}/t$ (first term), while at late times $t \gg t_{\rm d,0}^{\rm n}$ the dominant loss term is absorption by ejecta walls $\sim (1-A)E_{\rm nth}/t_{\rm d}^{\rm n}$ (last term).  For simplicity, we take $A = 0$ for our discussion here, although this does not affect the generality of our results since a non-zero value of $A$ can be absorbed into a redefinition of $t_{\rm d}^{\rm n}$.  Balancing source with losses in equation~(\ref{eq:photon_evo2}) thus gives an expression for the non-thermal energy
\begin{eqnarray}
E_{\rm nth} = \left\{
\begin{array}{lr}
L_{\rm sd}t = L_{\rm sd}t_{\rm d}^{\rm n}\left(\frac{t}{t_{\rm d,0}^{\rm n}}\right)^{3/2}, 
&  t \ll t_{\rm d,0}^{\rm n} \\
L_{\rm sd}t_{\rm d}^{\rm n}
,& t \gg t_{\rm d,0}^{\rm n}, \\
\end{array}
\right.
\label{eq:Enteq}
\end{eqnarray}
where in the top line we have used equation (\ref{eq:tdncases}).

Likewise, the thermal energy of the ejecta evolves according to (eq.~[\ref{eq:thermo_evo}])
\be
\frac{dE_{\rm th}}{dt} = -\frac{E_{\rm th}}{R}\frac{dR}{dt}  - L_{\rm th} + (1-A)\frac{E_{\rm nth}}{t_{\rm d}^{\rm n}}
\label{eq:thermo_evoeq}
\ee
Here the only source term is the absorption by the ejecta walls $\sim E_{\rm nth}/t_{\rm d}^{\rm n}$ (last term).  However, the dominant loss term changes with time.  At early times ($t \ll t_{\rm d,0}^{\rm ej}$) the dominant loss is PdV work $\sim E_{\rm th}/t$ (first term), while at late times ($t \gg t_{\rm d,0}^{\rm ej}$) the dominant loss term is radiation $L_{\rm th} \sim E_{\rm th}/t_{\rm d}^{\rm ej}$ (eq.~[\ref{eq:L_SN}]).  Balancing sources with losses and using equation (\ref{eq:tejcases}), we have
\begin{eqnarray}
\frac{E_{\rm th}}{t_{\rm d}^{\rm ej}} = \left\{
\begin{array}{lr}
\frac{E_{\rm th}}{t_{\rm d}^{\rm n}}\left(\frac{t}{t_{\rm d}^{\rm ej}}\right) = \frac{E_{\rm nth}}{t_{\rm d}^{\rm n}}\left(\frac{t}{t_{\rm d,0}^{\rm ej}}\right)^{2}, 
& t \ll t_{\rm d,0}^{\rm ej} \\
\frac{E_{\rm nth}}{t_{\rm d}^{\rm n}}
,& t \gg t_{\rm d,0}^{\rm ej} \\
\end{array}
\right.
\label{eq:Etheq}
\end{eqnarray}
where in the top line we have used equation (\ref{eq:tdncases}).

\subsection{Thermal Luminosity}

Using equation (\ref{eq:Etheq}) the thermal light curve $L_{\rm th} = E_{\rm th}/t_{\rm d}^{\rm ej}$ for the two cases is thus approximately given by
 \begin{eqnarray}
\frac{L_{\rm th}}{L_{\rm sd}} = \frac{E_{\rm th}}{t_{\rm d}^{\rm ej}L_{\rm sd}} = \left\{
\begin{array}{lr}
\chi^{-3/2}\left(\frac{t}{t_{\rm d,0}^{\rm ej}}\right)^{7/2}, 
& t \ll t_{\rm d,0}^{\rm ej}, t_{\rm d,0}^{\rm n} \\
\left(\frac{t}{t_{\rm d,0}^{\rm n}}\right)^{3/2}
,& t_{\rm d,0}^{\rm ej} \ll t \ll t_{\rm d,0}^{\rm n} \\
1
,& t \gg t_{\rm d,0}^{\rm n}, t_{\rm d,0}^{\rm ej} \\
\end{array}
\right.  (\chi > 1)
\label{eq:Lth1}
\end{eqnarray}
and 
 \begin{eqnarray}
\frac{L_{\rm th}}{L_{\rm sd}} = \frac{E_{\rm th}}{t_{\rm d}^{\rm ej}L_{\rm sd}} = \left\{
\begin{array}{lr}
\chi^{-3/2}\left(\frac{t}{t_{\rm d,0}^{\rm ej}}\right)^{7/2}, 
& t \ll t_{\rm d,0}^{\rm n}, t_{\rm d,0}^{\rm ej} \\
\left(\frac{t}{t_{\rm d,0}^{\rm ej}}\right)^{2}
,& t_{\rm d,0}^{\rm n} \ll t \ll t_{\rm d,0}^{\rm ej} \\
1
,& t \gg t_{\rm d,0}^{\rm n}, t_{\rm d,0}^{\rm ej} \\
\end{array}
\right.   (\chi < 1)
\label{eq:Lth2}
\end{eqnarray}
Since $L_p \propto t^{-2}$, the light curve first rises as $L_{\rm th} \propto t^{3/2}$, before flattening to $L_{\rm th} \propto t^{-1/2}$ or $\propto t^{0}$ in the $\chi >1$ and $\chi < 1$ cases, respectively.  Finally at late times the light curve tracks energy input from the pulsar $L_{\rm th} \propto L_{\rm sd} \propto t^{-2}$.  The light curve thus peaks at a time $t_{\rm peak} =$ min $[t_{\rm d,0}^{\rm n}, t_{\rm d,0}^{\rm ej}]$.  Since the ejecta expands linearly with time $R \propto t$, the effective temperature of the thermal radiation scales as $T_{\rm eff} \equiv (L_{\rm th}/4\pi \sigma R^{2})^{1/4} \propto L_{\rm th}^{1/4}t^{-1/2}$.

\begin{figure}
\includegraphics[width=0.5\textwidth]{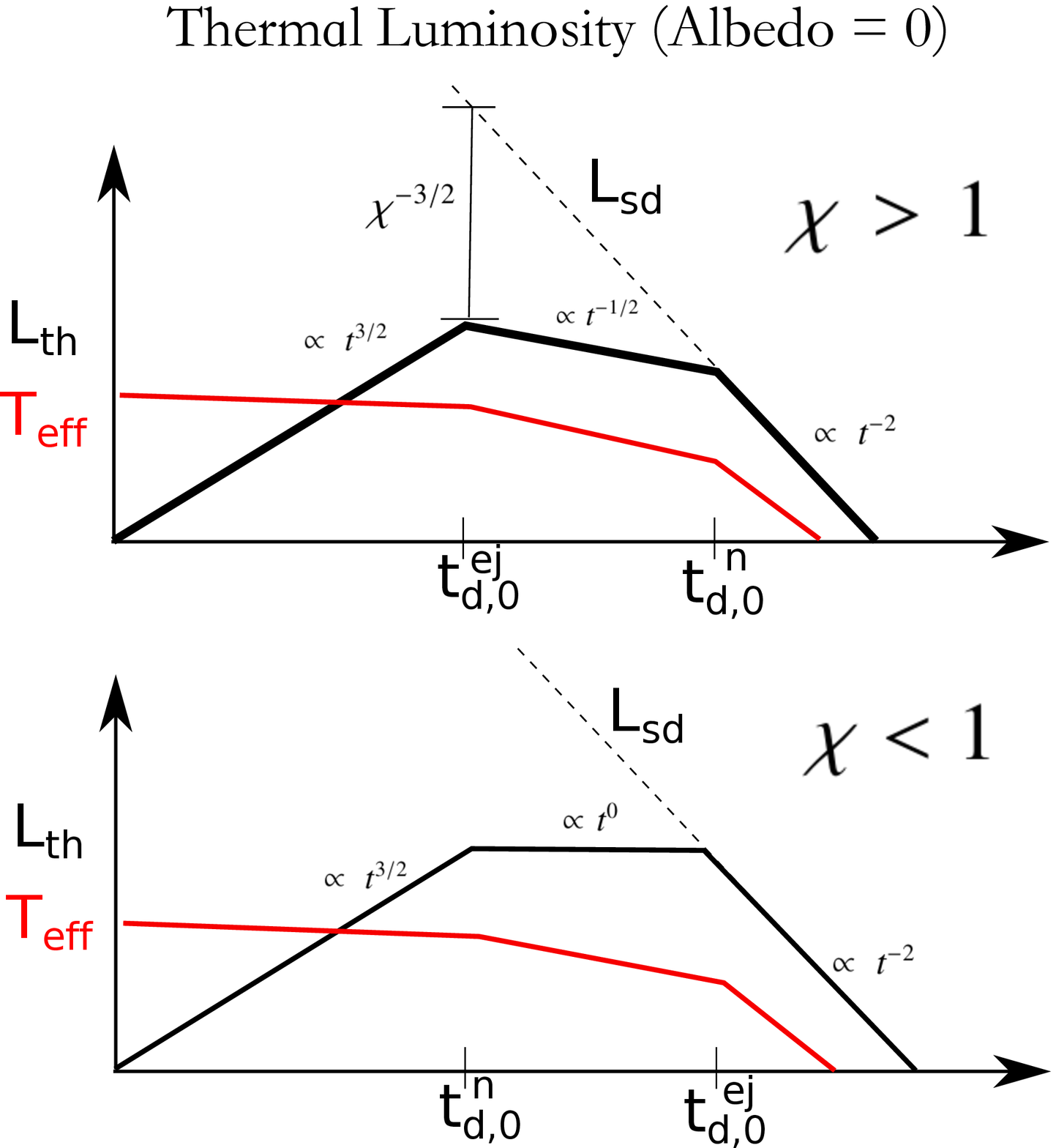}
\caption{Schematic diagram of the time evolution of thermal luminosity (eq.~[\ref{eq:Lth1}], [\ref{eq:Lth2}]) in the idealized case of perfectly absorbing ejecta (albedo = 0).  Shown for comparison with a dashed line is the pulsar luminosity $L_{\rm sd}$.  The top and bottom panels show cases corresponding to different values of the critical parameter $\chi \equiv t_{\rm d,0}^{\rm n}/t_{\rm d,0}^{\rm ej}$ (eq.~[\ref{eq:chi}]).  When $\chi > 1$ the peak luminosity is significantly lower than that of the pulsar.  This is due to the lower efficiency with which nebular X-rays thermalize as the result of PdV losses experienced due to the high pair optical depth through the nebula.   } 
\label{fig:schematicLC}
\end{figure}

Figure \ref{fig:schematicLC} shows a schematic diagram illustrating the evolution of $L_{\rm th}$ and $T_{\rm eff}$ in the $\chi > 1$ ({\it top}) and $\chi < 1$ ({\it bottom}) cases.  The pulsar luminosity $L_{\rm sd} \propto t^{-2}$ is shown for comparison.  In the $\chi > 1$ case the light curve peaks at $t = t_{\rm d,0}^{\rm ej}$ at a luminosity
\begin{eqnarray}
L_{\rm th,peak} &=& \nonumber \\
&&\left\{
\begin{array}{lr}

L_{\rm sd}|_{t = t_{\rm d,0}^{\rm ej}} \chi^{-3/2} \simeq 1.1\times 10^{45}B_{15}^{-1}M_{-2}^{7/4}\beta^{-1/4}  {\rm erg\,s^{-1}} , 
& \chi > 1 \\
L_{\rm sd}|_{t = t_{\rm d,0}^{\rm ej}} \simeq 6\times 10^{45}B_{15}^{-2}M_{-2}^{-1}\beta {\rm\,\, erg\,s^{-1}} 
,& \chi < 1 \\
\end{array}
\right.\nonumber \\
\label{eq:Lpeak}
\end{eqnarray}
In the $\chi < 1$ case, $L_{\rm th,peak}$ is close to that predicted by models which include only diffusion through the ejecta (\citealt{Kasen&Bildsten10}; Yu et al.~2010), as in the original model of \citet{Arnett82}.  In the \mbox{$\chi >1$} case, however, $L_{\rm th,peak}$ is suppressed by a factor $\chi^{-3/2}$ as compared to the case without pairs.  As already discussed in the main text, the physical reason for this suppression is the high optical depth of the pairs, which reduces the non-thermal energy density of the nebula because X-rays lose energy to PdV work on a timescale $\sim t$ faster than they can cross the nebula to be absorbed by the ejecta walls when $t \gg t_{\rm d,0}^{\rm n}$.

\begin{figure}
\includegraphics[width=0.5\textwidth]{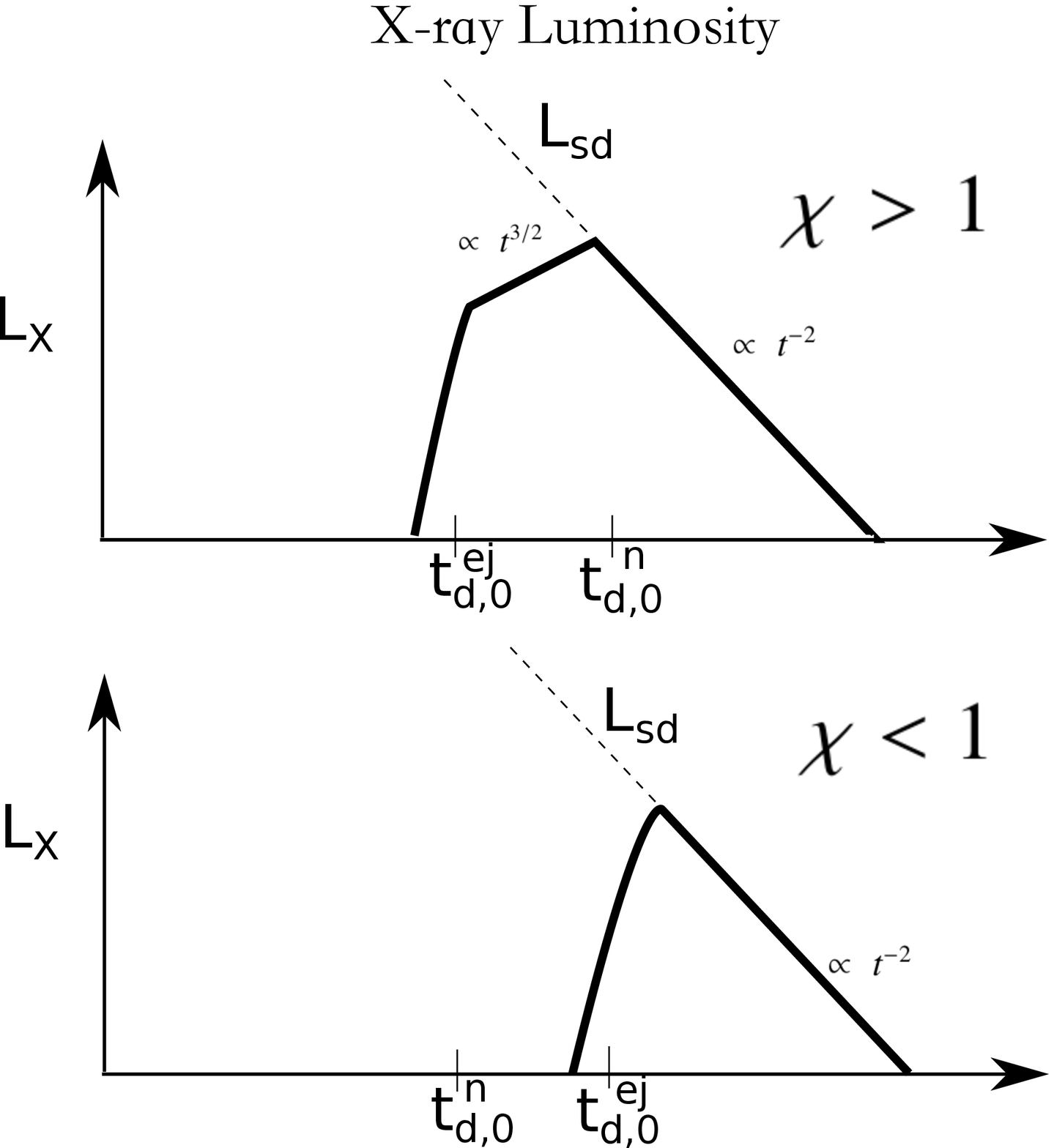}
\caption{Same as Figure \ref{fig:schematicLC}, but for the total non-thermal X-ray luminosity.  The X-ray luminosity in a given spectral band (e.g.~1-10 keV) is a factor $\approx 6$ times lower than the total luminosity.   } 
\label{fig:schematicLCX}
\end{figure}

\begin{figure}
\includegraphics[width=0.5\textwidth]{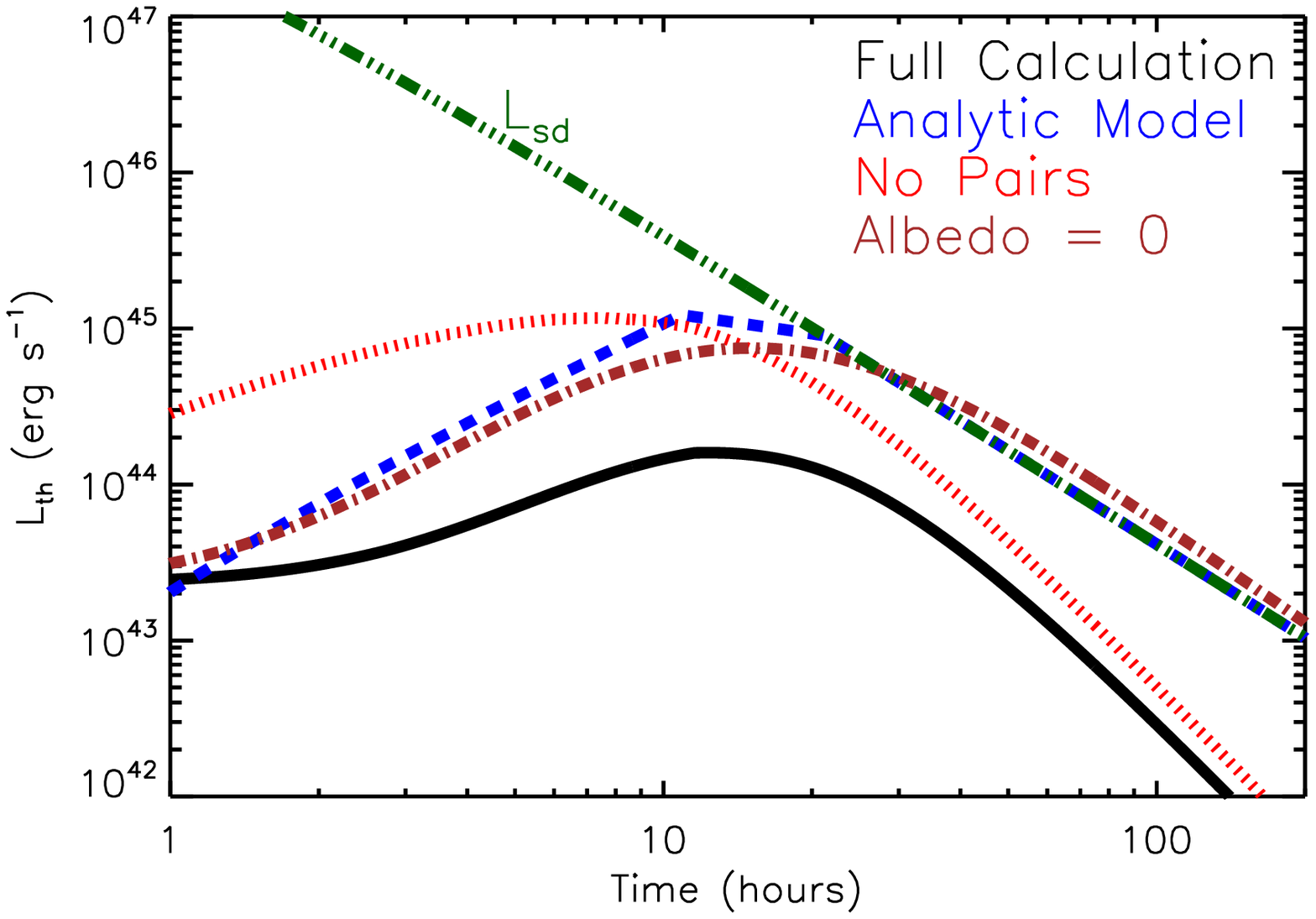}
\caption{Thermal luminosity $L_{\rm th}$ as a function of time for our fiducial model ($B_{\rm d} = 10^{15}$ G, $P = 2$ ms; $\chi = 2.7$), calculated under different assumptions, as compared to the pulsar luminosity ({\it green}).  The solid black line shows the actual result of our full calculation (Figure \ref{fig:evo15}).  The dotted red line shows a calculation in which we have artificially set the optical depth of pairs in the nebula to zero, an assumption implicitly made by previous models  (e.g.,~Kasen \& Bildsten 2010; Yu et al.~2013).  The dot-dashed brown line shows a calculation in which we have artificially set the albedo $\mathcal{A}_{\nu}$ equal to zero (pure absorption) at all frequencies.  The blue-dashed line shows our simple analytic model (eqs.~[\ref{eq:Lth1}], [\ref{eq:Lth2}]; Fig.~\ref{fig:schematicLC}), which was also derived under the $\mathcal{A}_{\nu} = 0$ assumption.} 
\label{fig:LCs}
\end{figure}

Equations (\ref{eq:Lth1}) and (\ref{eq:Lth2}) represent the {\it maximum} achievable luminosity of thermal emission since we have assumed perfect absorption ($A = 0$) in our analytic estimate.  Our full calculation predicts a lower luminosity due to the effects of finite albedo and the escape of radiation once the ejecta becomes optically thin.  These points are illustrated by Figure \ref{fig:LCs}, which compares the thermal light curve for our fiducial model ($B_{\rm d} = 10^{15}$; $M_{\rm ej} = 10^{-2}M_{\odot}$; Figs.~\ref{fig:evo15}, \ref{fig:ion15}) calculated under different assumptions.  The luminosity given by our full calculation ({\it black solid line}) is suppressed a factor of $\sim 10$ as compared to the simple theoretical estimate in equation (\ref{eq:Lpeak}) ({\it dashed blue line}) due to effects of finite albedo.  A similar suppression results due to the presence of nebular pairs, as can be seen by comparing to the otherwise identical model that artificially sets the pair optical depth to zero ({\it dotted red line}).    

\subsection{Non-Thermal X-ray Luminosity}

Unlike the thermal optical emission, X-rays can only escape the nebula once the ejecta becomes fully ionized.  As we discuss in Appendix \ref{sec:ion}, complete ionization is in fact readily achieved in many cases.  If the ejecta becomes ionized at time $t \sim t_{\rm d,0}^{\rm ej}$, then the {\it total} X-ray luminosity evolves approximately according to 
 \begin{eqnarray}
\frac{L_{\rm X,tot}}{L_{\rm sd}} = \frac{E_{\rm nth}}{6t_{d}^{\rm n}L_{\rm sd}} = \left\{
\begin{array}{lr}
\\
0, 
& t \ll t_{\rm d,0}^{\rm ej}, t_{\rm d,0}^{\rm n} \\
\left(\frac{t}{t_{\rm d,0}^{\rm n}}\right)^{3/2}
,& t_{\rm d,0}^{\rm ej} \ll t \ll t_{\rm d,0}^{\rm n} \\
1
,& t \gg t_{\rm d,0}^{\rm n}, t_{\rm d,0}^{\rm ej} \\
\end{array}
\right.  (\chi > 1)
\label{eq:Lnth1}
\end{eqnarray}\begin{eqnarray}
\frac{L_{\rm X,tot}}{L_{\rm sd}} = \frac{E_{\rm nth}}{t_{d}^{\rm n}L_{\rm sd}} = \left\{
\begin{array}{lr}
0, 
& t \ll t_{\rm d,0}^{\rm ej} \\
1
,& t \gg t_{\rm d,0}^{\rm ej} \\
\end{array}
\right.  (\chi < 1)
\label{eq:Lnth2}
\end{eqnarray}
where we have used equation (\ref{eq:Enteq}) for $E_{\rm nth}$.  Figure \ref{fig:schematicLCX} shows a schematic diagram illustrating the evolution of $L_{\rm X, tot}$ in the $\chi > 1$ ({\it top}) and $\chi < 1$ ({\it bottom}) cases.  

To obtain the X-ray luminosity in a given observing band (e.g.~0.1-1 keV or 1-10 keV) one should divide $L_{\rm X,tot}$ by factor of $\approx$ 6 to correct for the fraction of the non-thermal radiation $J_{\nu} \propto \nu^{-1}$ within a decade of X-ray energy.  When ionization break-out is achieved, the observed X-ray luminosity in a given band thus peaks on a timescale $t_{\rm peak,X} =$ max[$t_{\rm d,0}^{\rm ej}$, $t_{\rm d,0}^{\rm n}$] at a luminosity    
\begin{eqnarray}
L_{\rm X,peak}= \left\{
\begin{array}{lr}
L_{\rm sd}|_{t = t_{\rm d,0}^{\rm n}}/6 \simeq 1.4\times 10^{44}B_{15}^{-2/3}\beta^{2/3}  {\rm erg\,s^{-1}} , 
& \chi > 1 \\
L_{\rm sd}|_{t = t_{\rm d,0}^{\rm ej}}/6 \simeq 1\times 10^{45}B_{15}^{-2}M_{-2}^{-1}\beta {\rm\,\, erg\,s^{-1}} 
,& \chi < 1 \\
\end{array}
\right.\nonumber \\
\label{eq:LpeakX}
\end{eqnarray}

\section{Ionization State of Ejecta}
\label{sec:ion}

This section provides analytic estimates of the conditions required to fully ionize the merger ejecta, as is required for the escape of X-ray emission.  As discussed in $\S\ref{sec:ejecta}$, we assume that the ejecta is composed of iron, for which the ionization threshold energy of the innermost (hydrogen-like) electronic state is $h \nu_{\rm thr} \simeq 9.1\,{\rm keV}$.

The characteristic penetration depth $\Delta$ of a photon of frequency $\nu > \nu_{\rm thr}$ is achieved at an effective optical depth of unity (eq.~[\ref{eq:tau1}]), as defined by the integral condition
\be
\int_{0}^{s = \Delta}\rho_{\rm sh}\kappa_{\rm abs,\nu}(s')\left[1 + \kappa_{\rm es}\rho_{\rm sh}s'\right]ds' = 1,
\label{eq:tau2}
\ee
where $s$ is the depth through the shocked layer and $\kappa_{\rm abs,\nu}(s) = f_n(s)\sigma_{\rm bf,\nu}/56 m_p$ is the bound-free opacity.  The radiative cross section for hydrogen-like species is approximately given by (e.g.,~\citealt{Osterbrock&Ferland06})
\be
\sigma_{\rm bf}(\nu) = 1.2\times 10^{-20}\left(\frac{\nu_{\rm thr}}{\nu}\right)^{3}{\rm cm^{2}},
\label{eq:sigmabf}
\ee
The neutral fraction $f_{\rm n} \ll 1$ (eq.~[\ref{eq:fn}]) 
\begin{eqnarray}
f_n(s) &\approx& \frac{\alpha_{\rm rec} n_e V}{c}\left(\int_{\nu_{\rm min}(s)}^{\infty}\frac{E_{\rm nth,\nu}}{h\nu}\sigma_{\rm bf}(\nu)d\nu\right)^{-1},
\label{eq:fna}
\end{eqnarray}
varies with depth $s$ because only photons that have not yet been absorbed by this depth. This is because those with $\nu > \nu_{\rm min}(s)$ contribute to the local photo-ionization rate, whereas $\nu_{\rm min}$ is the frequency determined by condition given by equation (\ref{eq:tau2}) for $\Delta = s$.  Here $\alpha_{\rm rec}$ is the recombination coefficient, which is approximately given by (e.g.,~\citealt{Nahar+01})
\be
\alpha_{\rm rec} = 2\times 10^{-10}T_{5}^{-0.6}{\rm cm^{3}s^{-1}},
\label{eq:alpharec}
\ee
where $T_{\rm C} \simeq 10^{5}T_{5}K$ is the ejecta temperature, normalized to a value characteristic of the Compton temperature of electrons in the ionized layer near the time of ionization break-out.

For photons at the ionization threshold frequency $\nu_{\rm thr}$, the penetration depth can be derived from equation (\ref{eq:tau2}) assuming a constant opacity $\kappa_{\rm abs,\nu}$ with depth, calculated using the unattenuated nebular spectrum (i.e. $\nu_{\rm min} = \nu_{\rm thr}$ in equation \ref{eq:fna}):
\be
\frac{\Delta}{\Delta_{\rm sh}} \simeq \frac{\sqrt{1 + 4 \eta_{\rm thr}^{-1}} - 1}{2 \tau_{\rm es}^{\rm ej}} \underset{\eta_{\rm th} \gg 1}\approx \frac{1}{\tau_{\rm es}^{\rm ej}\eta_{\rm th}},
\label{eq:dth}
\ee
where $\Delta_{\rm sh}$ and $\tau_{\rm es}^{\rm ej} = \rho_{\rm ej}\kappa_{\rm es}\Delta_{\rm sh}$ are the total thickness and scattering optical depth of the shocked ejecta (eq.~[\ref{eq:tau_ej}]), respectively, and
\begin{eqnarray}
\eta_{\rm thr} &\equiv& \left.\frac{\kappa_{\rm abs,\nu}}{\kappa_{\rm es}}\right|_{\nu = \nu_{\rm thr}} \simeq \frac{\alpha_{\rm rec}n_{e}Vh \nu_{\rm thr}}{14 \kappa_{\rm es} m_p c E_{\rm nth,\nu_{\rm thr}}\nu_{\rm thr}} \approx \frac{\alpha_{\rm rec}M_{\rm ej} h \nu_{\rm thr} (1 - \mathcal{A}_{\nu})}{2\kappa_{\rm es} m_p^{2} c\beta L_{\rm sd}t } \nonumber \\
&\approx& 17(1-\mathcal{A}_{\nu})\left(\frac{L_{\rm sd} t}{10^{50}{\rm erg}}\right)^{-1}M_{-2}\beta^{-1}T_{5}^{-0.6} \gg 1
\label{eq:etaA}
\end{eqnarray}
is the ratio of absorptive and scattering opacity at $\nu = \nu_{\rm thr}$, where in equation (\ref{eq:etaA}) we estimate the electron density as $n_e \approx \rho_{\rm ej}/2m_p \sim M_{\rm ej}/2m_p V$ (eq.~[\ref{eq:rho0}]) and calculate the ionization rate assuming $E_{\rm nth,\nu} \propto \nu^{-1}$.  The photon energy of the nebula $E_{\rm nth,\nu_{\rm thr}}$ at frequency $\nu_{\rm thr}$ is approximated by assuming a balance between energy injection from the pulsar and losses due to absorption by the ejecta walls (as was employed in deriving equation [\ref{eq:Enteq}])
\be
\dot{E}_{\rm sd,\nu} \approx (1 - \mathcal{A}_{\nu})\frac{E_{\rm nth,\nu}}{t_{\rm d}^{\rm n}} \,\,\,\,\Rightarrow\,\,\,\, E_{\rm nth, \nu_{\rm thr}}\nu_{\rm thr} \approx \frac{L_{\rm sd}t}{14(1-\mathcal{A}_{\nu})}\frac{v}{c}
\label{eq:edotsdeq}
\ee
where $\mathcal{A}_{\nu}$ is the albedo of the ionized layer at $\nu_{\rm thr}$, $\dot{E}_{\rm sd,\nu_{\rm thr}}\nu_{\rm thr} \approx L_{\rm sd}/14$ is the rate of energy injection by the pulsar (eq.~[\ref{eq:edotsd}]), and we have approximated $t_{\rm d}^{\rm n} \sim R/c $ (eq.~[\ref{eq:tdiffn}]) as the photon crossing time of the nebula (formally this is valid near the optical peak only if $\chi < 1$, but it is sufficient for purposes of an estimate).  

 From equation (\ref{eq:dth}), the condition for ionization break-out ($\Delta \gtrsim \Delta_{\rm sh}$) can be written as a condition on the scattering optical depth of the shocked ejecta $\tau_{\rm es}^{\rm ej}$ (eq.~[\ref{eq:tau_ej}]):
\begin{eqnarray}
 \tau_{\rm es}^{\rm ej} < \tau_{\rm es,bo}^{\rm ej} = 2\eta_{\rm thr}^{-1/2}
\end{eqnarray}
where the pre-factor of $2$ accounts for the somewhat deeper penetration depth of photons with $\nu \gtrsim 2\nu_{\rm thr}$  (\citealt{Metzger+13}) than those at the ionization threshold frequency,\footnote{
Equation (\ref{eq:dth}) provides the penetration depth of photons at the threshold frequency $\nu_{\rm thr}$, but higher energy photons penetrate somewhat further due to the smaller absorption cross section $\sigma_{\rm bf,\nu} \propto \nu^{-3}$.  \citet{Metzger+13} show that photons with $\nu \sim 2\nu_{\rm thr}$ thus penetrate to a distance which is $\approx 2-3$ times larger than those at the threshold frequency, while $\Delta$ increases much slower with frequency for $\nu \gg \nu_{\rm thr}$, due to the diminishing number of ionizing photons at higher frequency $E_{\rm nth,\nu} \propto \nu^{-1}$.} the former chosen to represent a `typical' range of frequencies in the band of interest.  

The break-out time corresponding to this optical depth is 
\begin{eqnarray}
t_{\rm bo} = \left(\frac{3M_{\rm ej}\kappa_{\rm es}}{4\pi v_{\rm ej}^{2}\tau_{\rm es,bo}^{\rm ej}}\right)^{1/2} \approx 10.8\,{\rm hr}\,M_{-2}^{3/4}\beta^{-5/4}T_{5}^{-0.15}\left(\frac{L_{\rm sd}t}{10^{50}{\rm ergs}}\right)^{-1/4}
\label{eq:tbo}
\end{eqnarray}
where we have adopted a fiducial value of the albedo $\mathcal{A}_{\nu} \sim 0.5$ and have used equation (\ref{eq:etaA}) for $\eta_{\rm thr}$ and equation (\ref{eq:tau_ej}) for $\tau^{\rm ej}_{\rm es}$.  

Now substituting in the pulsar luminosity (eq.~[\ref{eq:Lsd}]) at times $t \ll t_{\rm sd}$, we can solve for the break-out time explicitly
\begin{eqnarray}
t_{\rm bo} \approx  8.4\,M_{-2}^{2}\beta^{-3}T_{5}^{-0.6}B_{15}^{2} \,{\rm hr} \approx 8.4\,M_{-2}^{7/2}T_{5}^{-0.6}B_{15}^{2} \,{\rm hr},   
\label{eq:tbo2}    
\end{eqnarray}
where in the last line we have again made use of the approximation relationship between ejecta mass and velocity assuming a total rotational energy $E_{\rm rot} = 10^{52}$ ergs.  Equation (\ref{eq:tbo2}) shows that the break-out time is a strongly increasing function of the ejecta mass and the pulsar magnetic field.  If $M_{\rm ej}$ and/or $B_{\rm d}$ are sufficiently high, then complete ionization is delayed to times much greater than the optical peak luminosity, preventing the escape of non-thermal X-rays from the nebula.  These trends are apparent in our numerical models shown in Figure \ref{fig:models_lum}.


\begin{thebibliography}{}

\bibitem[\protect\citeauthoryear{{Abadie}, {Abbott}, {Abbott}, {Abernathy},
  {Accadia}, {Acernese}, {Adams}, {Adhikari}, {Ajith}, {Allen} \& et
  al.}{{Abadie} et~al.}{2010}]{Abadie+10}
{Abadie} J.,  {Abbott} B.~P.,  {Abbott} R.,  {Abernathy} M.,  {Accadia} T.,
  {Acernese} F.,  {Adams} C.,  {Adhikari} R.,  {Ajith} P.,  {Allen} B.,    et
  al. 2010, Classical and Quantum Gravity, 27, 173001

\bibitem[\protect\citeauthoryear{{Abdikamalov}, {Ott}, {Rezzolla}, {Dessart},
  {Dimmelmeier}, {Marek} \& {Janka}}{{Abdikamalov}
  et~al.}{2010}]{Abdikamalov+10}
{Abdikamalov} E.~B.,  {Ott} C.~D.,  {Rezzolla} L.,  {Dessart} L.,
  {Dimmelmeier} H.,  {Marek} A.,    {Janka} H.-T.,  2010, \prd, 81, 044012

\bibitem[\protect\citeauthoryear{{Accadia} \& {Virgo Collaboration}}{{Accadia}
  \& {Virgo Collaboration}}{2011}]{VIRGO}
{Accadia} T.,  {Virgo Collaboration} 2011, Classical and Quantum Gravity, 28,
  114002

\bibitem[\protect\citeauthoryear{{Antoniadis} et~al.,}{{Antoniadis}
  et~al.}{2013}]{Antoniadis+13}
{Antoniadis} J.,  et~al., 2013, Science, 340, 448

\bibitem[\protect\citeauthoryear{{Arnett}}{{Arnett}}{1982}]{Arnett82}
{Arnett} W.~D.,  1982, \apj, 253, 785

\bibitem[\protect\citeauthoryear{{Barnes} \& {Kasen}}{{Barnes} \&
  {Kasen}}{2013}]{Barnes&Kasen13}
{Barnes} J.,  {Kasen} D.,  2013, ApJ, submitted, arXiv:1303.5787

\bibitem[\protect\citeauthoryear{{Belczynski}, {O'Shaughnessy}, {Kalogera},
  {Rasio}, {Taam} \& {Bulik}}{{Belczynski} et~al.}{2008}]{Belczynski+08}
{Belczynski} K.,  {O'Shaughnessy} R.,  {Kalogera} V.,  {Rasio} F.,  {Taam}
  R.~E.,    {Bulik} T.,  2008, \apjl, 680, L129

\bibitem[\protect\citeauthoryear{{Berger}, {Fong} \& {Chornock}}{{Berger}
  et~al.}{2013}]{Berger+13}
{Berger} E.,  {Fong} W.,    {Chornock} R.,  2013, \apjl, 774, L23

\bibitem[\protect\citeauthoryear{{Bostanc{\i}}, {Kaneko} \& {G{\"o}{\u
  g}{\"u}{\c s}}}{{Bostanc{\i}} et~al.}{2013}]{Bostanci+13}
{Bostanc{\i}} Z.~F.,  {Kaneko} Y.,    {G{\"o}{\u g}{\"u}{\c s}} E.,  2013,
  \mnras, 428, 1623

\bibitem[\protect\citeauthoryear{{Bromberg}, {Nakar}, {Piran} \&
  {Sari}}{{Bromberg} et~al.}{2011}]{Bromberg+11}
{Bromberg} O.,  {Nakar} E.,  {Piran} T.,    {Sari} R.,  2011, \apj, 740, 100

\bibitem[\protect\citeauthoryear{{Bucciantini}, {Metzger}, {Thompson} \&
  {Quataert}}{{Bucciantini} et~al.}{2012}]{Bucciantini+12}
{Bucciantini} N.,  {Metzger} B.~D.,  {Thompson} T.~A.,    {Quataert} E.,  2012,
  \mnras, 419, 1537

\bibitem[\protect\citeauthoryear{{Caballero}, {McLaughlin} \&
  {Surman}}{{Caballero} et~al.}{2012}]{Caballero+12}
{Caballero} O.~L.,  {McLaughlin} G.~C.,    {Surman} R.,  2012, \apj, 745, 170

\bibitem[\protect\citeauthoryear{{Cenko} et~al.,}{{Cenko}
  et~al.}{2013}]{Cenko+13}
{Cenko} S.~B.,  et~al., 2013, \apj, 769, 130

\bibitem[\protect\citeauthoryear{{Chawla}, {Anderson}, {Besselman}, {Lehner},
  {Liebling}, {Motl} \& {Neilsen}}{{Chawla} et~al.}{2010}]{Chawla+10}
{Chawla} S.,  {Anderson} M.,  {Besselman} M.,  {Lehner} L.,  {Liebling} S.~L.,
  {Motl} P.~M.,    {Neilsen} D.,  2010, Physical Review Letters, 105, 111101

\bibitem[\protect\citeauthoryear{{Cucchiara}, {Prochaska}, {Perley}, {Cenko},
  {Werk}, {Cao}, {Bloom} \& {Cobb}}{{Cucchiara} et~al.}{2013}]{Cucchiara+13}
{Cucchiara} A.,  {Prochaska} J.~X.,  {Perley} D.~A.,  {Cenko} S.~B.,  {Werk}
  J.,  {Cao} Y.,  {Bloom} J.~S.,    {Cobb} B.~E.,  2013, ArXiv e-prints

\bibitem[\protect\citeauthoryear{{Dall'Osso}, {Stratta}, {Guetta}, {Covino},
  {De Cesare} \& {Stella}}{{Dall'Osso} et~al.}{2011}]{DallOsso+11}
{Dall'Osso} S.,  {Stratta} G.,  {Guetta} D.,  {Covino} S.,  {De Cesare} G.,
  {Stella} L.,  2011, \aap, 526, A121

\bibitem[\protect\citeauthoryear{{Darbha}, {Metzger}, {Quataert}, {Kasen},
  {Nugent} \& {Thomas}}{{Darbha} et~al.}{2010}]{Darbha+10}
{Darbha} S.,  {Metzger} B.~D.,  {Quataert} E.,  {Kasen} D.,  {Nugent} P.,
  {Thomas} R.,  2010, \mnras, pp 1254--+

\bibitem[\protect\citeauthoryear{{Demorest}, {Pennucci}, {Ransom}, {Roberts} \&
  {Hessels}}{{Demorest} et~al.}{2010}]{Demorest+10}
{Demorest} P.~B.,  {Pennucci} T.,  {Ransom} S.~M.,  {Roberts} M.~S.~E.,
  {Hessels} J.~W.~T.,  2010, \nat, 467, 1081

\bibitem[\protect\citeauthoryear{{Dessart}, {Burrows}, {Ott}, {Livne}, {Yoon}
  \& {Langer}}{{Dessart} et~al.}{2006}]{Dessart+06}
{Dessart} L.,  {Burrows} A.,  {Ott} C.~D.,  {Livne} E.,  {Yoon} S.,    {Langer}
  N.,  2006, \apj, 644, 1063

\bibitem[\protect\citeauthoryear{{Dessart}, {Ott}, {Burrows}, {Rosswog} \&
  {Livne}}{{Dessart} et~al.}{2009}]{Dessart+09}
{Dessart} L.,  {Ott} C.~D.,  {Burrows} A.,  {Rosswog} S.,    {Livne} E.,  2009,
  \apj, 690, 1681

\bibitem[\protect\citeauthoryear{{Duncan} \& {Thompson}}{{Duncan} \&
  {Thompson}}{1992}]{Duncan&Thompson92}
{Duncan} R.~C.,  {Thompson} C.,  1992, \apjl, 392, L9

\bibitem[\protect\citeauthoryear{{Eichler}, {Livio}, {Piran} \&
  {Schramm}}{{Eichler} et~al.}{1989}]{Eichler+89}
{Eichler} D.,  {Livio} M.,  {Piran} T.,    {Schramm} D.~N.,  1989, \nat, 340,
  126

\bibitem[\protect\citeauthoryear{{Fern{\'a}ndez} \& {Metzger}}{{Fern{\'a}ndez}
  \& {Metzger}}{2013}]{Fernandez&Metzger13}
{Fern{\'a}ndez} R.,  {Metzger} B.~D.,  2013, ArXiv e-prints

\bibitem[\protect\citeauthoryear{{Fong} \& {Berger}}{{Fong} \&
  {Berger}}{2013}]{Fong&Berger13}
{Fong} W.,  {Berger} E.,  2013, \apj, 776, 18

\bibitem[\protect\citeauthoryear{{Fong}, {Berger}, {Metzger}, {Margutti},
  {Chornock}, {Migliori}, {Foley}, {Zauderer}, {Lunnan}, {Laskar}, {Desch},
  {Meech}, {Sonnett}, {Dickey}, {Hedlund} \& {Harding}}{{Fong}
  et~al.}{2013}]{Fong+13}
{Fong} W.-f.,  {Berger} E.,  {Metzger} B.~D.,  {Margutti} R.,  {Chornock} R.,
  {Migliori} G.,  {Foley} R.~J.,  {Zauderer} B.~A.,  {Lunnan} R.,  {Laskar} T.,
   {Desch} S.~J.,  {Meech} K.~J.,  {Sonnett} S.,  {Dickey} C.~M.,  {Hedlund}
  A.~M.,    {Harding} P.,  2013, ArXiv e-prints

\bibitem[\protect\citeauthoryear{{Freiburghaus}, {Rosswog} \&
  {Thielemann}}{{Freiburghaus} et~al.}{1999}]{Freiburghaus+99}
{Freiburghaus} C.,  {Rosswog} S.,    {Thielemann} F.,  1999, \apjl, 525, L121

\bibitem[\protect\citeauthoryear{{Gao}, {Ding}, {Wu}, {Zhang} \& {Dai}}{{Gao}
  et~al.}{2013}]{Gao+13}
{Gao} H.,  {Ding} X.,  {Wu} X.-F.,  {Zhang} B.,    {Dai} Z.-G.,  2013, \apj,
  771, 86

\bibitem[\protect\citeauthoryear{{Gao} \& {Fan}}{{Gao} \&
  {Fan}}{2006}]{Gao&Fan06}
{Gao} W.-H.,  {Fan} Y.-Z.,  2006, Chinese Journal of Astronomy \& Astrophysics,
  6, 513

\bibitem[\protect\citeauthoryear{{Giacomazzo} \& {Perna}}{{Giacomazzo} \&
  {Perna}}{2013}]{Giacomazzo&Perna13}
{Giacomazzo} B.,  {Perna} R.,  2013, \apjl, 771, L26

\bibitem[\protect\citeauthoryear{{Gompertz}, {O'Brien}, {Wynn} \&
  {Rowlinson}}{{Gompertz} et~al.}{2013}]{Gompertz+13}
{Gompertz} B.~P.,  {O'Brien} P.~T.,  {Wynn} G.~A.,    {Rowlinson} A.,  2013,
  \mnras, 431, 1745

\bibitem[\protect\citeauthoryear{{Harry} \& {LIGO Scientific
  Collaboration}}{{Harry} \& {LIGO Scientific Collaboration}}{2010}]{LIGO}
{Harry} G.~M.,  {LIGO Scientific Collaboration} 2010, Classical and Quantum
  Gravity, 27, 084006

\bibitem[\protect\citeauthoryear{{Hebeler}, {Lattimer}, {Pethick} \&
  {Schwenk}}{{Hebeler} et~al.}{2013}]{Hebeler+13}
{Hebeler} K.,  {Lattimer} J.~M.,  {Pethick} C.~J.,    {Schwenk} A.,  2013,
  \apj, 773, 11

\bibitem[\protect\citeauthoryear{{Hotokezaka}, {Kiuchi}, {Kyutoku}, {Okawa},
  {Sekiguchi}, {Shibata} \& {Taniguchi}}{{Hotokezaka}
  et~al.}{2013}]{Hotokezaka+13}
{Hotokezaka} K.,  {Kiuchi} K.,  {Kyutoku} K.,  {Okawa} H.,  {Sekiguchi} Y.-i.,
  {Shibata} M.,    {Taniguchi} K.,  2013, Phys. Rev. D, 87, 024001

\bibitem[\protect\citeauthoryear{{Hotokezaka}, {Kyutoku}, {Okawa}, {Shibata} \&
  {Kiuchi}}{{Hotokezaka} et~al.}{2011}]{Hotokezaka+11}
{Hotokezaka} K.,  {Kyutoku} K.,  {Okawa} H.,  {Shibata} M.,    {Kiuchi} K.,
  2011, \prd, 83, 124008

\bibitem[\protect\citeauthoryear{{Kaplan}, {Ott}, {O'Connor}, {Kiuchi},
  {Roberts} \& {Duez}}{{Kaplan} et~al.}{2013}]{Kaplan+13}
{Kaplan} J.~D.,  {Ott} C.~D.,  {O'Connor} E.~P.,  {Kiuchi} K.,  {Roberts} L.,
   {Duez} M.,  2013, ArXiv e-prints

\bibitem[\protect\citeauthoryear{{Kasen}, {Badnell} \& {Barnes}}{{Kasen}
  et~al.}{2013}]{Kasen+13}
{Kasen} D.,  {Badnell} N.~R.,    {Barnes} J.,  2013, ApJ, submitted,
  arXiv:1303.5788

\bibitem[\protect\citeauthoryear{{Kasen} \& {Bildsten}}{{Kasen} \&
  {Bildsten}}{2010}]{Kasen&Bildsten10}
{Kasen} D.,  {Bildsten} L.,  2010, \apj, 717, 245

\bibitem[\protect\citeauthoryear{{Kasliwal} \& {Nissanke}}{{Kasliwal} \&
  {Nissanke}}{2013}]{Kasliwal&Nissanke13}
{Kasliwal} M.~M.,  {Nissanke} S.,  2013, ArXiv e-prints

\bibitem[\protect\citeauthoryear{{Kennel} \& {Coroniti}}{{Kennel} \&
  {Coroniti}}{1984}]{Kennel&Coroniti84}
{Kennel} C.~F.,  {Coroniti} F.~V.,  1984, \apj, 283, 694

\bibitem[\protect\citeauthoryear{{Kiziltan}, {Kottas}, {De Yoreo} \&
  {Thorsett}}{{Kiziltan} et~al.}{2013}]{Kiziltan+13}
{Kiziltan} B.,  {Kottas} A.,  {De Yoreo} M.,    {Thorsett} S.~E.,  2013, ArXiv
  e-prints

\bibitem[\protect\citeauthoryear{{Law} et~al.,}{{Law}  et~al.}{2009}]{Law+09}
{Law} N.~M.,  et~al., 2009, \pasp, 121, 1395

\bibitem[\protect\citeauthoryear{{Lee}, {Ramirez-Ruiz} \&
  {L{\'o}pez-C{\'a}mara}}{{Lee} et~al.}{2009}]{Lee+09}
{Lee} W.~H.,  {Ramirez-Ruiz} E.,    {L{\'o}pez-C{\'a}mara} D.,  2009, \apjl,
  699, L93

\bibitem[\protect\citeauthoryear{{Lehner}, {Palenzuela}, {Liebling}, {Thompson}
  \& {Hanna}}{{Lehner} et~al.}{2012}]{Lehner+12}
{Lehner} L.,  {Palenzuela} C.,  {Liebling} S.~L.,  {Thompson} C.,    {Hanna}
  C.,  2012, PRD, 86, 104035

\bibitem[\protect\citeauthoryear{{Li} \& {Paczy{\'n}ski}}{{Li} \&
  {Paczy{\'n}ski}}{1998}]{Li&Paczynski98}
{Li} L.,  {Paczy{\'n}ski} B.,  1998, ApJ, 507, L59

\bibitem[\protect\citeauthoryear{{Metzger} \& {Berger}}{{Metzger} \&
  {Berger}}{2012}]{Metzger&Berger12}
{Metzger} B.~D.,  {Berger} E.,  2012, \apj, 746, 48

\bibitem[\protect\citeauthoryear{{Metzger} \& {Bower}}{{Metzger} \&
  {Bower}}{2013}]{Metzger&Bower13}
{Metzger} B.~D.,  {Bower} G.~C.,  2013, ArXiv e-prints

\bibitem[\protect\citeauthoryear{{Metzger}, {Giannios} \& {Mimica}}{{Metzger}
  et~al.}{2012}]{Metzger+12}
{Metzger} B.~D.,  {Giannios} D.,    {Mimica} P.,  2012, \mnras, 420, 3528

\bibitem[\protect\citeauthoryear{{Metzger}, {Giannios}, {Thompson},
  {Bucciantini} \& {Quataert}}{{Metzger} et~al.}{2011}]{Metzger+11}
{Metzger} B.~D.,  {Giannios} D.,  {Thompson} T.~A.,  {Bucciantini} N.,
  {Quataert} E.,  2011, \mnras, 413, 2031

\bibitem[\protect\citeauthoryear{{Metzger}, {Mart{\'{\i}}nez-Pinedo}, {Darbha},
  {Quataert}, {Arcones}, {Kasen}, {Thomas}, {Nugent}, {Panov} \&
  {Zinner}}{{Metzger} et~al.}{2010a}]{Metzger+10}
{Metzger} B.~D.,  {Mart{\'{\i}}nez-Pinedo} G.,  {Darbha} S.,  {Quataert} E.,
  {Arcones} A.,  {Kasen} D.,  {Thomas} R.,  {Nugent} P.,  {Panov} I.~V.,
  {Zinner} N.~T.,  2010a, \mnras, 406, 2650

\bibitem[\protect\citeauthoryear{{Metzger}, {Mart{\'{\i}}nez-Pinedo}, {Darbha},
  {Quataert}, {Arcones}, {Kasen}, {Thomas}, {Nugent}, {Panov} \&
  {Zinner}}{{Metzger} et~al.}{2010b}]{Metzger+10b}
{Metzger} B.~D.,  {Mart{\'{\i}}nez-Pinedo} G.,  {Darbha} S.,  {Quataert} E.,
  {Arcones} A.,  {Kasen} D.,  {Thomas} R.,  {Nugent} P.,  {Panov} I.~V.,
  {Zinner} N.~T.,  2010b, \mnras, 406, 2650

\bibitem[\protect\citeauthoryear{{Metzger}, {Piro} \& {Quataert}}{{Metzger}
  et~al.}{2009a}]{Metzger+09a}
{Metzger} B.~D.,  {Piro} A.~L.,    {Quataert} E.,  2009a, \mnras, 396, 304

\bibitem[\protect\citeauthoryear{{Metzger}, {Piro} \& {Quataert}}{{Metzger}
  et~al.}{2009b}]{Metzger+09b}
{Metzger} B.~D.,  {Piro} A.~L.,    {Quataert} E.,  2009b, \mnras, 396, 1659

\bibitem[\protect\citeauthoryear{{Metzger}, {Quataert} \& {Thompson}}{{Metzger}
  et~al.}{2008}]{Metzger+08}
{Metzger} B.~D.,  {Quataert} E.,    {Thompson} T.~A.,  2008, \mnras, 385, 1455

\bibitem[\protect\citeauthoryear{{Metzger}, {Thompson} \& {Quataert}}{{Metzger}
  et~al.}{2007}]{Metzger+07}
{Metzger} B.~D.,  {Thompson} T.~A.,    {Quataert} E.,  2007, \apj, 659, 561

\bibitem[\protect\citeauthoryear{{Metzger}, {Thompson} \& {Quataert}}{{Metzger}
  et~al.}{2008}]{Metzger+08b}
{Metzger} B.~D.,  {Thompson} T.~A.,    {Quataert} E.,  2008, \apj, 676, 1130

\bibitem[\protect\citeauthoryear{{Metzger}, {Vurm}, {Hascoet} \&
  {Beloborodov}}{{Metzger} et~al.}{2013}]{Metzger+13}
{Metzger} B.~D.,  {Vurm} I.,  {Hascoet} R.,    {Beloborodov} A.~M.,  2013,
  ArXiv e-prints

\bibitem[\protect\citeauthoryear{{Morrison}, {Baumgarte} \&
  {Shapiro}}{{Morrison} et~al.}{2004}]{Morrison+04}
{Morrison} I.~A.,  {Baumgarte} T.~W.,    {Shapiro} S.~L.,  2004, \apj, 610, 941

\bibitem[\protect\citeauthoryear{{Nahar}}{{Nahar}}{2006}]{Nahar06}
{Nahar} S.~N.,  2006, \aap, 457, 721

\bibitem[\protect\citeauthoryear{{Nahar}, {Eissner}, {Chen} \&
  {Pradhan}}{{Nahar} et~al.}{2009}]{Nahar+09}
{Nahar} S.~N.,  {Eissner} W.,  {Chen} G.~X.,    {Pradhan} A.~K.,  2009, VizieR
  Online Data Catalog, 340, 80789

\bibitem[\protect\citeauthoryear{{Nahar} \& {Pradhan}}{{Nahar} \&
  {Pradhan}}{1994}]{Nahar&Pradhan94}
{Nahar} S.~N.,  {Pradhan} A.~K.,  1994, \pra, 49, 1816

\bibitem[\protect\citeauthoryear{{Nahar}, {Pradhan}, {Chen} \&
  {Eissner}}{{Nahar} et~al.}{2011}]{Nahar+11}
{Nahar} S.~N.,  {Pradhan} A.~K.,  {Chen} G.-X.,    {Eissner} W.,  2011, \pra,
  83, 053417

\bibitem[\protect\citeauthoryear{{Nahar}, {Pradhan} \& {Zhang}}{{Nahar}
  et~al.}{2001}]{Nahar+01}
{Nahar} S.~N.,  {Pradhan} A.~K.,    {Zhang} H.~L.,  2001, \apjs, 133, 255

\bibitem[\protect\citeauthoryear{{Narayan}, {Paczynski} \& {Piran}}{{Narayan}
  et~al.}{1992}]{Narayan+92}
{Narayan} R.,  {Paczynski} B.,    {Piran} T.,  1992, \apjl, 395, L83

\bibitem[\protect\citeauthoryear{{Nissanke}, {Kasliwal} \&
  {Georgieva}}{{Nissanke} et~al.}{2013}]{Nissanke+13}
{Nissanke} S.,  {Kasliwal} M.,    {Georgieva} A.,  2013, ApJ, 767, 124

\bibitem[\protect\citeauthoryear{{Nomoto} \& {Kondo}}{{Nomoto} \&
  {Kondo}}{1991}]{Nomoto&Kondo91}
{Nomoto} K.,  {Kondo} Y.,  1991, \apjl, 367, L19

\bibitem[\protect\citeauthoryear{{Norris} \& {Bonnell}}{{Norris} \&
  {Bonnell}}{2006}]{Norris&Bonnell06}
{Norris} J.~P.,  {Bonnell} J.~T.,  2006, \apj, 643, 266

\bibitem[\protect\citeauthoryear{{O'Connor} \& {Ott}}{{O'Connor} \&
  {Ott}}{2011}]{OConnor&Ott11}
{O'Connor} E.,  {Ott} C.~D.,  2011, \apj, 730, 70

\bibitem[\protect\citeauthoryear{{Oechslin} \& {Janka}}{{Oechslin} \&
  {Janka}}{2006}]{Oechslin&Janka06}
{Oechslin} R.,  {Janka} H.-T.,  2006, \mnras, 368, 1489

\bibitem[\protect\citeauthoryear{{Osterbrock} \& {Ferland}}{{Osterbrock} \&
  {Ferland}}{2006}]{Osterbrock&Ferland06}
{Osterbrock} D.~E.,  {Ferland} G.~J.,  2006, {Astrophysics of gaseous nebulae
  and active galactic nuclei}

\bibitem[\protect\citeauthoryear{{{\"O}zel}, {Psaltis}, {Ransom}, {Demorest} \&
  {Alford}}{{{\"O}zel} et~al.}{2010}]{Ozel+10}
{{\"O}zel} F.,  {Psaltis} D.,  {Ransom} S.,  {Demorest} P.,    {Alford} M.,
  2010, \apjl, 724, L199

\bibitem[\protect\citeauthoryear{{Paczynski}}{{Paczynski}}{1986}]{Paczynski86}
{Paczynski} B.,  1986, \apjl, 308, L43

\bibitem[\protect\citeauthoryear{{Paschalidis}, {Etienne} \&
  {Shapiro}}{{Paschalidis} et~al.}{2012}]{Paschalidis+12}
{Paschalidis} V.,  {Etienne} Z.~B.,    {Shapiro} S.~L.,  2012, \prd, 86, 064032

\bibitem[\protect\citeauthoryear{{Perley} et~al.,}{{Perley}
  et~al.}{2009}]{Perley+09}
{Perley} D.~A.,  et~al., 2009, \apj, 696, 1871

\bibitem[\protect\citeauthoryear{{Pinto} \& {Eastman}}{{Pinto} \&
  {Eastman}}{2000}]{Pinto&Eastman00}
{Pinto} P.~A.,  {Eastman} R.~G.,  2000, \apj, 530, 757

\bibitem[\protect\citeauthoryear{{Piran}, {Nakar} \& {Rosswog}}{{Piran}
  et~al.}{2013}]{Piran+13}
{Piran} T.,  {Nakar} E.,    {Rosswog} S.,  2013, MNRAS, 430, 2121

\bibitem[\protect\citeauthoryear{{Piro}}{{Piro}}{2008}]{Piro08}
{Piro} A.~L.,  2008, \apj, 679, 616

\bibitem[\protect\citeauthoryear{{Piro} \& {Kulkarni}}{{Piro} \&
  {Kulkarni}}{2013}]{Piro&Kulkarni13}
{Piro} A.~L.,  {Kulkarni} S.~R.,  2013, \apjl, 762, L17

\bibitem[\protect\citeauthoryear{{Piro} \& {Ott}}{{Piro} \&
  {Ott}}{2011}]{Piro&Ott11}
{Piro} A.~L.,  {Ott} C.~D.,  2011, \apj, 736, 108

\bibitem[\protect\citeauthoryear{{Porth}, {Komissarov} \& {Keppens}}{{Porth}
  et~al.}{2013}]{Porth+13}
{Porth} O.,  {Komissarov} S.~S.,    {Keppens} R.,  2013, \mnras, 431, L48

\bibitem[\protect\citeauthoryear{{Price} \& {Rosswog}}{{Price} \&
  {Rosswog}}{2006}]{Price&Rosswog06}
{Price} D.~J.,  {Rosswog} S.,  2006, Science, 312, 719

\bibitem[\protect\citeauthoryear{{Rezzolla}, {Baiotti}, {Giacomazzo}, {Link} \&
  {Font}}{{Rezzolla} et~al.}{2010}]{Rezzolla+10}
{Rezzolla} L.,  {Baiotti} L.,  {Giacomazzo} B.,  {Link} D.,    {Font} J.~A.,
  2010, Classical and Quantum Gravity, 27, 114105

\bibitem[\protect\citeauthoryear{{Rezzolla}, {Giacomazzo}, {Baiotti}, {Granot},
  {Kouveliotou} \& {Aloy}}{{Rezzolla} et~al.}{2011}]{Rezzolla+11}
{Rezzolla} L.,  {Giacomazzo} B.,  {Baiotti} L.,  {Granot} J.,  {Kouveliotou}
  C.,    {Aloy} M.~A.,  2011, \apjl, 732, L6

\bibitem[\protect\citeauthoryear{{Roberts}, {Kasen}, {Lee} \&
  {Ramirez-Ruiz}}{{Roberts} et~al.}{2011}]{Roberts+11}
{Roberts} L.~F.,  {Kasen} D.,  {Lee} W.~H.,    {Ramirez-Ruiz} E.,  2011, \apjl,
  736, L21+

\bibitem[\protect\citeauthoryear{{Rosswog}, {Korobkin}, {Arcones} \&
  {Thielemann}}{{Rosswog} et~al.}{2013}]{Rosswog+13b}
{Rosswog} S.,  {Korobkin} O.,  {Arcones} A.,    {Thielemann} F.-K.,  2013,
  ArXiv e-prints

\bibitem[\protect\citeauthoryear{{Rosswog} \& {Liebend{\"o}rfer}}{{Rosswog} \&
  {Liebend{\"o}rfer}}{2003}]{Rosswog&Liebendorfer03}
{Rosswog} S.,  {Liebend{\"o}rfer} M.,  2003, \mnras, 342, 673

\bibitem[\protect\citeauthoryear{{Rosswog}, {Piran} \& {Nakar}}{{Rosswog}
  et~al.}{2012}]{Rosswog+12}
{Rosswog} S.,  {Piran} T.,    {Nakar} E.,  2012, ArXiv e-prints

\bibitem[\protect\citeauthoryear{{Rowlinson}, {O'Brien}, {Metzger}, {Tanvir} \&
  {Levan}}{{Rowlinson} et~al.}{2013}]{Rowlinson+13}
{Rowlinson} A.,  {O'Brien} P.~T.,  {Metzger} B.~D.,  {Tanvir} N.~R.,    {Levan}
  A.~J.,  2013, \mnras, 430, 1061

\bibitem[\protect\citeauthoryear{Ruffert \& Janka}{Ruffert \&
  Janka}{1999}]{ruffert1999}
Ruffert M.,  Janka H.-T.,  1999, A\&A, 344, 573

\bibitem[\protect\citeauthoryear{{Shappee} et~al.,}{{Shappee}
  et~al.}{2013}]{Shappee+13}
{Shappee} B.~J.,  et~al., 2013, ArXiv e-prints

\bibitem[\protect\citeauthoryear{{Somiya}}{{Somiya}}{2012}]{KAGRA}
{Somiya} K.,  2012, Classical and Quantum Gravity, 29, 124007

\bibitem[\protect\citeauthoryear{{Svensson}}{{Svensson}}{1987}]{Svensson87}
{Svensson} R.,  1987, \mnras, 227, 403

\bibitem[\protect\citeauthoryear{{Tanaka} \& {Hotokezaka}}{{Tanaka} \&
  {Hotokezaka}}{2013}]{Tanaka&Hotokezaka13}
{Tanaka} M.,  {Hotokezaka} K.,  2013, \apj, 775, 113

\bibitem[\protect\citeauthoryear{{Tanaka}, {Hotokezaka}, {Kyutoku}, {Wanajo},
  {Kiuchi}, {Sekiguchi} \& {Shibata}}{{Tanaka} et~al.}{2013}]{Tanaka+13}
{Tanaka} M.,  {Hotokezaka} K.,  {Kyutoku} K.,  {Wanajo} S.,  {Kiuchi} K.,
  {Sekiguchi} Y.,    {Shibata} M.,  2013, ArXiv e-prints

\bibitem[\protect\citeauthoryear{{Tanvir}, {Levan}, {Fruchter}, {Hjorth},
  {Hounsell}, {Wiersema} \& {Tunnicliffe}}{{Tanvir} et~al.}{2013}]{Tanvir+13}
{Tanvir} N.~R.,  {Levan} A.~J.,  {Fruchter} A.~S.,  {Hjorth} J.,  {Hounsell}
  R.~A.,  {Wiersema} K.,    {Tunnicliffe} R.~L.,  2013, \nat, 500, 547

\bibitem[\protect\citeauthoryear{{Tauris}, {Sanyal}, {Yoon} \&
  {Langer}}{{Tauris} et~al.}{2013}]{Tauris+13}
{Tauris} T.~M.,  {Sanyal} D.,  {Yoon} S.-C.,    {Langer} N.,  2013, \aap, 558,
  A39

\bibitem[\protect\citeauthoryear{{Thompson} \& {Duncan}}{{Thompson} \&
  {Duncan}}{1995}]{Thompson&Duncan95}
{Thompson} C.,  {Duncan} R.~C.,  1995, in {Fruchter} A.~S.,  {Tavani} M.,
  {Backer} D.~C.,  eds, Millisecond Pulsars. A Decade of Surprise Vol.~72 of
  Astronomical Society of the Pacific Conference Series, {Accretion-Induced
  Collapse and Neutron Star Magnetic Fields}.
p.~301

\bibitem[\protect\citeauthoryear{{Thompson}, {Chang} \& {Quataert}}{{Thompson}
  et~al.}{2004}]{Thompson+04}
{Thompson} T.~A.,  {Chang} P.,    {Quataert} E.,  2004, \apj, 611, 380

\bibitem[\protect\citeauthoryear{{Ury{\= u}}, {Shibata} \& {Eriguchi}}{{Ury{\=
  u}} et~al.}{2000}]{Uryu+00}
{Ury{\= u}} K.~{\= o}.,  {Shibata} M.,    {Eriguchi} Y.,  2000, \prd, 62,
  104015

\bibitem[\protect\citeauthoryear{{Usov}}{{Usov}}{1992}]{Usov92}
{Usov} V.~V.,  1992, \nat, 357, 472

\bibitem[\protect\citeauthoryear{{Wu}, {Gao}, {Ding}, {Zhang}, {Dai} \&
  {Wei}}{{Wu} et~al.}{2013}]{Wu+13}
{Wu} X.-F.,  {Gao} H.,  {Ding} X.,  {Zhang} B.,  {Dai} Z.-G.,    {Wei} J.-Y.,
  2013, ArXiv e-prints

\bibitem[\protect\citeauthoryear{{Yu}, {Zhang} \& {Gao}}{{Yu}
  et~al.}{2013}]{Yu+13}
{Yu} Y.-W.,  {Zhang} B.,    {Gao} H.,  2013, ArXiv e-prints

\bibitem[\protect\citeauthoryear{{Zhang}}{{Zhang}}{2013}]{Zhang13}
{Zhang} B.,  2013, \apjl, 763, L22

\bibitem[\protect\citeauthoryear{{Zrake} \& {MacFadyen}}{{Zrake} \&
  {MacFadyen}}{2013}]{Zrake&MacFadyen13}
{Zrake} J.,  {MacFadyen} A.~I.,  2013, \apjl, 769, L29

\end{thebibliography}


\end{document}